\documentclass[nofootinbib,reprint,amsmath,amssymb,aps]{revtex4-1}
\usepackage{graphicx}
\usepackage{dcolumn}
\usepackage{bm}
\usepackage{hyperref}
\usepackage{comment}

\newcommand\RS{{\mathrm{R_\star}}}

\begin{document}

\preprint{APS/123-QED}

\title[Three-dimensional general relativistic Poynting-Robertson effect II]{Three-dimensional general relativistic Poynting-Robertson effect II:\\ Radiation field from a rigidly rotating spherical source}

\author{Pavel Bakala$^{1,2,5}$}\email{pavel.bakala@physics.cz}
\author{Vittorio De Falco$^{1}$},\email{vittorio.defalco@physics.cz}
\author{Emmanuele Battista$^{1}$}\email{emmanuelebattista@gmail.com}
\author{Kate\v{r}ina Goluchov\'a$^{1}$}
\author{Debora Lan\v{c}ov\'a$^{1}$}
\author{Maurizio Falanga$^{3,4}$} 
\author{Luigi Stella$^{5}$\vspace{0.5cm}}

\affiliation{$^1$ Research Centre for Computational Physics and Data Processing, Faculty of Philosophy \& Science, Silesian University in Opava, Bezru\v{c}ovo n\'am.~13, CZ-746\,01 Opava, Czech Republic \\
$^2$M. R. \v{S}tef\'anik Observatory and Planetarium, Sl\'adkovi\v{c}ova 41, 920 01 Hlohovec, Slovak Republic\\
$^3$ International Space Science Institute, Hallerstrasse 6, 3012 Bern, Switzerland\\
$^4$ International Space Science Institute Beijing, No.1 Nanertiao, Zhongguancun, Haidian District, 100190 Beijing, China\\
$^5$ INAF -  Osservatorio Astronomico di Roma,  Via Frascati, 33, Monteporzio Catone, 00078  Roma, Italy}

\date{\today}

\begin{abstract}
We investigate the three-dimensional, general relativistic Poynting-Robertson (PR) effect in the case of 
rigidly rotating spherical source which emits radiation radially in the local comoving frame. 
Such radiation field is meant to approximate the field produced by the surface of a rotating neutron star, or 
by the central radiating hot corona of accreting black holes; it extends the purely radial radiation 
field that we considered in a previous study. Its angular momentum is expressed in terms
of the rotation frequency and radius of the emitting source. For the background we adopt a Kerr spacetime geometry.
We derive the equations of motion for test particles influenced by such radiation field, recovering the 
classical and weak-field approximation for slow rotation. We concentrate on solutions consisting of particles
orbiting along circular orbits off and parallel to the equatorial plane,  which are stabilized by the balance between gravitational attraction, radiation force and PR drag. Such solutions are found to lie on a critical hypersurface, whose shape may morph from prolate to oblate depending on the Kerr spin parameter and the luminosity, rotation and radius of the radiating sphere. For selected parameter ranges, the critical hypersurface intersects the radiating sphere giving rise to a bulging equatorial region or, alternatively, two lobes above the poles. We calculate the trajectories of test particles in the close vicinity of the critical hypersurface for a selected set of initial parameters and analyze the spatial and angular velocity of test particles captured on the critical hypersurface. 
\end{abstract}

\maketitle
\section{Introduction}
\label{sec:intro}

Radiation emitted from a central massive body has long been 
known to affect the orbit of gravitating particles both as an result of the 
radial, outward-directed force and "radiation drag" caused by the 
so-called Poynting-Robertson (PR) effect 
\citep{Poynting1903,Robertson1937}. Originally studied in relation to 
the motion of dust, comets and meteors in the solar 
systems \cite{Wyatt1950,Guess1961,Burns1979},
the PR effect has been discussed in more recent years in the context of matter 
accreting towards compact objects, especially black holes (BHs) of 
all masses and neutron stars (NSs), which radiate at a sizable fraction 
of their Eddington luminosity or very close to it ({\it e.g.}\citep{Walker1992}). 
This has motivated extending the treatment of the PR effect to the more 
general framework of General Relativity (GR)\cite{Abramowicz1990,Lamb1995,Miller1996,Miller1998}. 
Early attempts at handling the non-geodesic (dissipative) PR equations of motion in GR   
were based on a \emph{direct spacetime approach} \cite{Abramowicz1988,Abramowicz1990,Prasanna1990,Iyer1993,Abramowicz1993}) and provided
a limited, though inspiring, description of the effect. Only in 2009--2011 Bini and his collaborators \cite{Bini2009,Bini2011} derived a full GR description of the PR effect by using a more suitable formalism (the \emph{observer-splitting formalism} \cite{Jantzen1992,Bini1997a,Bini1997b,Bini1998,Bini1999}). Recently, the Lagrangian formulation of the PR effect, involving an analytical form of the Rayleigh potential, has been developed \cite{Defalco2018,DeFalco2019,DeFalco2019VE}. Moreover, we have recently worked out the three-dimensional (3D) generalization of the two-dimensional model (2D) of Bini and collaborators \cite{DeFalco20183D}. The radiation field assumed in \cite{DeFalco20183D} consists of photons emitted in a purely radial direction with respect to the Zero Angular Momentum Observer (ZAMO) frame of the Kerr spacetime background. In that case a coherent set of dynamical equations is derived, based on which particle 
motion under the combined effect of radiation and gravity is determined and the 
critical hypersurface, where all forces attain equilibrium, is found. New effects such as the 
latitudinal drift of the trajectories and existence of suspended circular orbits were investigated. Based on these results, Wielgius performed a 3D treatment by employing finite size radiation effects framed in a simplified Kerr spacetime, including only linear terms in the spin \cite{Wielgus2019}. 

In the present work we adopt a different prescription for the 3D radiation field, which provides a somewhat more realistic description of emission properties of rotating compact 
objects or the hot coronae around them.  Its properties are designed to match at the equator 
those of the 2D radiation field introduced in \cite{Bini2011}.
The paper is structured as follows: in Sec. \ref{sec:geometry} we introduce the geometrical setup of our model. Sec. \ref{sec:rad_field} describes the radiation field, an approximation of that emitted from a rigidly rotating spherical surface. In Sec. \ref{sec:eoms} we derive the GR equations of motion 
for test particles in the Kerr spacetime under the influence of the radiation field. 
In Sec. \ref{sec:critc_rad} we discuss the salient features of the critical hypersurface on which radiation forces balance gravity, and present calculations of selected orbits in the vicinity of critical hypersurfaces.  Our conclusions are in Sec. \ref{sec:end}. 

\section{Geometric setup}
\label{sec:geometry}

\begin{figure*}[t]
\centering
\includegraphics[scale=0.6]{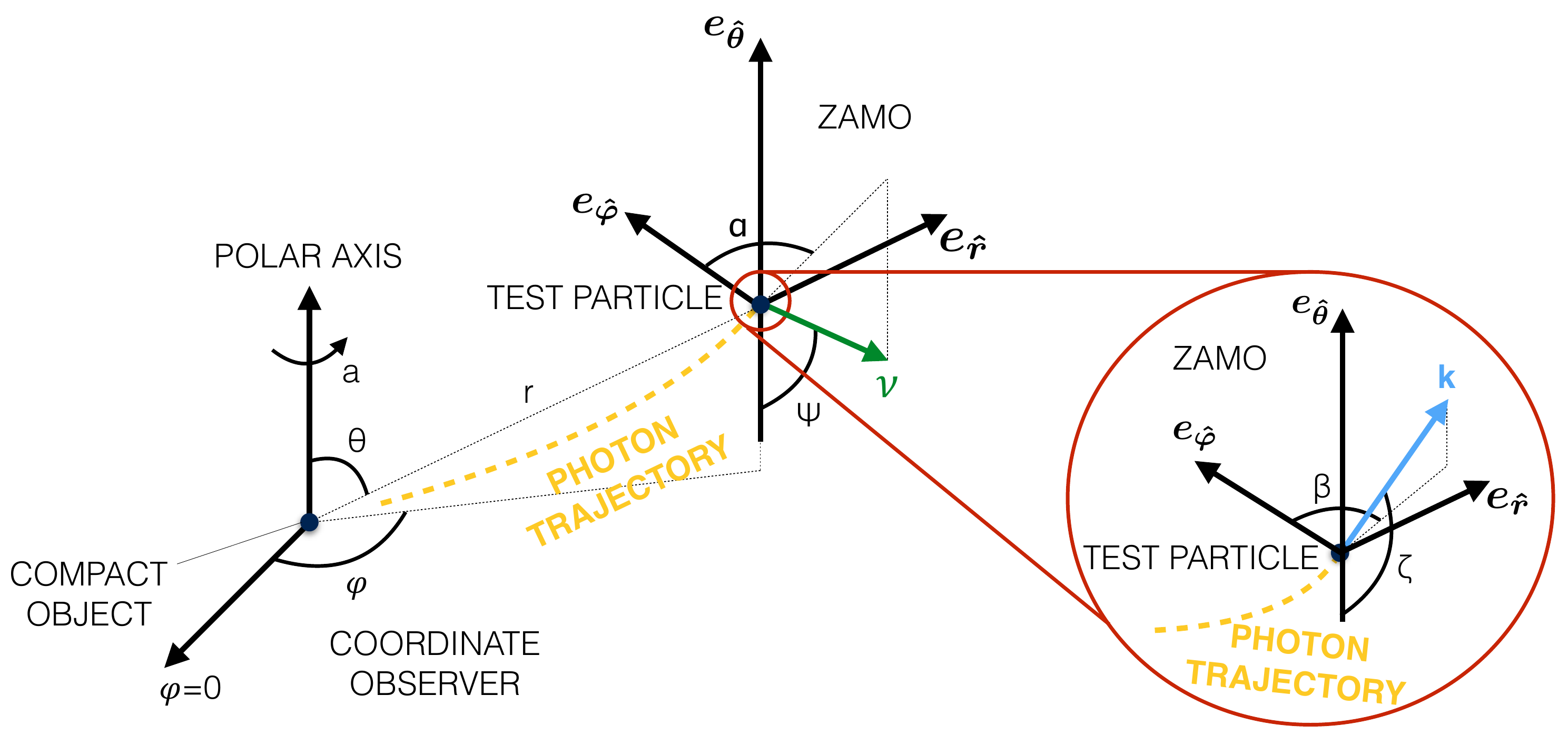}
\caption{Visual representation of the radiation field-test particle interaction geometry. The spatial location of the test particle is described by Boyer-Linquist coordinates $\left\{\boldsymbol{r},\boldsymbol{\theta},\boldsymbol{\varphi}\right\}$. The ZAMO local frame is $\left\{\boldsymbol{e_{\hat{t}}},\boldsymbol{e_{\hat{r}}},\boldsymbol{e_{\hat{\theta}}},\boldsymbol{e_{\hat{\varphi}}}\right\}$. The photons of the radiation field travel along null geodesics with four-momentum $\boldsymbol{k}$.  Two photon impact parameters, $b$ and $q$ are related respectively to the two angles $\beta$ and $\zeta$, measured in the local ZAMO frame. The test particle moves with a velocity described by the magnitude $\nu$ and two angles $\alpha$ and $\psi$, measured in the local ZAMO frame.}
\label{fig:Fig0}
\end{figure*}

We consider a central compact object, whose outer spacetime geometry is described by the Kerr metric. Using geometric units ($c = G = 1$) and metric signature $(-,+,+,+)$, the line element of the Kerr spacetime expressed in Boyer-Lindquist coordinates, parameterized by the mass $M$ and the Kerr parameter (spin) $a$, reads as \cite{Chandrasekhar1992}:
\begin{equation}\label{kerr_metric}
\begin{aligned}
 \mathrm{d}s^2 &= -\left(1-\frac{2Mr}{\Sigma}\right) \,\mathrm{d}t^2 
  - \frac{4Mra}{\Sigma} \sin^2\theta\,\mathrm{d}t\, \mathrm{d}\varphi \\
  &+ \frac{\Sigma}{\Delta}\, \mathrm{d}r^2 
  + \Sigma \,\mathrm{d}\theta^2
  + \rho\sin^2\theta\, \mathrm{d}\varphi^2, 
\end{aligned}
\end{equation}
where $\Sigma \equiv r^{2} + a^{2}\cos^{2}\theta$, $\Delta \equiv r^{2} - 2Mr + a^{2}$, and $\rho  \equiv r^2+a^2+2Ma^2r\sin^2\theta/\Sigma$. The determinant of the Kerr metric reads $g=-\Sigma^2\sin^{2}\theta$. In the Kerr spacetime the radial coordinates of the outer event horizon $R_{\rm +}$ and of the static limit (outer boundary of the ergosphere) $R_{\rm SL}$ are given by \cite{Misner1973}
\begin{equation}
\begin{aligned}
R_{\rm +}&=M(1+\sqrt{1-a^2}),\\
R_{\rm SL}&=M(1+\sqrt{1-a^2\cos^2\theta})\,,
\end{aligned}
\end{equation}
respectively. 

We consider the zero angular momentum observers (ZAMOs), whose future-pointing unit normal to the spatial hypersurfaces is given by
\begin{equation}\label{n}
\boldsymbol{n}=\frac{1}{N}(\boldsymbol{\partial_t}-N^{\varphi}\boldsymbol{\partial_\varphi})\,,
\end{equation}
where $N=(-g^{tt})^{-1/2}$ is the time lapse function and $N^{\varphi}=g_{t\varphi}/g_{\varphi\varphi}$ the spatial shift vector field. We conduct our investigations outside the event horizon, where the time coordinate hypersurfaces are spacelike, i.e., $g^{tt}<0$. An orthonormal frame adapted to the ZAMOs is 
\begin{equation} \label{eq:zamoframes}
\begin{aligned}
&\boldsymbol{e_{\hat t}}=\boldsymbol{n},\quad
\boldsymbol{e_{\hat r}}=\frac1{\sqrt{g_{rr}}}\boldsymbol{\partial_r},\\
&\boldsymbol{e_{\hat \theta}}=\frac1{\sqrt{g_{\theta \theta }}}\boldsymbol{\partial_\theta},\quad
\boldsymbol{e_{\hat \varphi}}=\frac1{\sqrt{g_{\varphi \varphi }}}\boldsymbol{\partial_\varphi}.
\end{aligned}
\end{equation}
All the vector and tensor indices (e.g., $v^\alpha, T^{\alpha\beta}$) associated to the ZAMO frame will be labeled by a hat (e.g., $v^{\hat\alpha}, T^{\hat{\alpha}\hat{\beta}}$), instead all the scalar quantities measured in the ZAMO frame (e.g., $f$) will be followed by $(n)$ (e.g., $f(n)$). In the kinematic decomposition of the ZAMO congruence, we have that the nonzero ZAMO kinematic quantities are acceleration $\boldsymbol{a}(n)=\nabla_{\boldsymbol{n}} \boldsymbol{n}$, expansion tensor along the $\hat{\varphi}$-direction $\boldsymbol{\theta_{\hat\varphi}}(n)$, and the relative Lie curvature vector $\boldsymbol{k_{(\rm Lie)}}(n)$ (see Table 1 in \cite{DeFalco20183D} for their explicit expressions and for further details).
 
\section{Radiation field}
\label{sec:rad_field}
We consider a radiation field consisting of a flux of photons traveling along null geodesics in the Kerr geometry. The related stress-energy tensor is \cite{DeFalco20183D}
\begin{equation}\label{STE}
T^{\mu\nu}=\Phi^2 k^\mu k^\nu\,,\qquad k^\mu k_\mu=0,\qquad k^\mu \nabla_\mu k^\nu=0,
\end{equation}
where $\Phi$ is a parameter linked to the radiation field intensity and $\boldsymbol{k}$ is the photon four-momentum field. By splitting $\boldsymbol{k}$ with respect to the ZAMO frame, we obtain
\begin{equation} \label{photon}
\begin{aligned}
&\boldsymbol{k}=E(n)[\boldsymbol{n}+\boldsymbol{\hat{\nu}}(k,n)],\\
&\boldsymbol{\hat{\nu}}(k,n)=\sin\zeta\sin\beta\ \boldsymbol{e_{\hat r}}+\cos\zeta\ \boldsymbol{e_{\hat\theta}}+\sin\zeta \cos\beta\ \boldsymbol{e_{\hat\varphi}},
\end{aligned}
\end{equation}
where $\boldsymbol{\hat{\nu}}(k,n)$ is the photon spatial unit relative velocity with respect to the ZAMOs, $\beta$ and $\zeta$ are the two angles measured in the ZAMO frame in the azimuthal and polar direction, respectively (see Fig. \ref{fig:Fig0}). The radiation field is governed by the two impact parameters $(b,q)$, associated respectively with the two emission angles $(\beta,\zeta)$. The photon energy, $E(n)$, and the photon angular momentum along the polar direction $\hat{\theta}$, $L_{\hat\theta}(n)$, with respect to the ZAMO frame is expressed in the frame of distant static observer by the following formulae \cite{DeFalco20183D}
\begin{eqnarray} \label{energyZAMO}
&&E(n)=\frac{E}{N}(1+bN^\varphi),\\
&&L_{\hat\theta}(n)\equiv E(n)\cos\beta\sin\zeta=\frac{L_z}{\sqrt{g_{\varphi\varphi}}},
\end{eqnarray}
where $E=-k_t>0$ is the conserved photon energy, $L_z=k_\varphi$ is the conserved angular momentum along the polar $z$ axis orthogonal to the equatorial plane, and $b=L_z/E$ is the azimuthal photon impact parameter, where all these quantities are measured by a distant static observer. At this impact parameter, we associate the relative azimuthal angle in the ZAMO frame \cite{DeFalco20183D}
\begin{equation} \label{ANG1}
\cos\beta=\frac{b E}{\sin\zeta\sqrt{g_{\varphi\varphi}}E(n)}=\frac{b N}{\sin\zeta\sqrt{g_{\varphi\varphi}}(1+b N^\varphi)}.
\end{equation}
The photon four-momentum components in the Kerr geometry are given by \cite{Chandrasekhar1992} 
\begin{equation} \label{CarterEQs} 
\begin{aligned}
     k^{t} &= \Sigma^{-1} \left(ab-a^2\sin^{2}\theta+(r^2+a^2)P\,\Delta^{-1}\right)\,,\\
		 k^{r} &=  s_r\Sigma^{-1} \sqrt{R_{b,q}(r)},\\
     k^{\theta} &=  s_{\theta}\Sigma^{-1} \sqrt{\Theta_{b,q}(\theta)}\,\\ 
		 k^{\varphi} &=  \Sigma^{-1}\left(b\,\mathrm{cosec}^{2} \theta-a+a\,P\,\Delta^{-1}\right),
\end{aligned}
\end{equation} 
where $P\equiv r^{2} + a^{2}-ba$, and the pair of signs $s_{r}$, $s_{\theta}$ describes the orientation of the radial and latitudinal evolution, respectively. The radial and latitudinal effective potentials are respectively \cite{Chandrasekhar1992}
\begin{equation}
\label{Rpot}
       R_{b,q} \left( r \right) = \left( r^{2} + a^{2}-ab \right)^{2}
       - \Delta \left[ q + \left( b - a \right) ^{2} \right], 
\end{equation} 			
\begin{equation}	
\label{Thetapot}			
\Theta_{b,q} \left( \theta \right) = q + a^{2} \cos^{2}
\theta -b^{2} \mathrm{cot}^{2} \theta,
\end{equation} 
where  
 \begin{equation} \label{q_def2}
q \equiv \left(\frac{k_{\theta}}{k_{t}}\right)^{2} + \left[b\tan\left(\frac{\pi}{2} - \theta\right)\right]^{2} - a^{2}\cos^{2}\theta\,,    
 \end{equation}
is the latitudinal photon impact parameter related to Carter's constant of motion $\mathcal{Q}$ through \cite{Chandrasekhar1992}
\begin{equation} \label{Carter_constant}
q=\mathcal{Q}/E^2.
 \end{equation}
\subsection{Radiation field from a spherical rigidly rotating surface}
\label{sec:rot_rad_field}
In the previous paper \cite{DeFalco20183D} we considered a test radiation field made by photons moving along purely radial directions in any ZAMO frame and even at infinity, independently of the radial coordinate. Here we consider a somewhat more complex model, consisting of a radiation field emitted from a spherical surface with a radius $\RS$ centered at the origin of the Boyer-Lindquist coordinates and rigidly rotating with angular velocity $\Omega_{\mathrm{\star}}$. The azimuthal motion of the photons, as measured by a static observer at infinity, is now determined not only by frame-dragging (as in \cite{DeFalco20183D}), but also by rotation of the surface. The present radiation field model is thus more general than our previous model.  

The four-momentum of a photon outgoing from a rigidly rotating spherical surface in a purely radial direction with respect to the corotating local observer frame labelled by $\left\{\boldsymbol{e_{\langle t \rangle}},\boldsymbol{e_{\langle r \rangle}},\boldsymbol{e_{\langle \theta \rangle}},\boldsymbol{e_{\langle \varphi \rangle}}\right\}$ is
\begin{equation} \label{local_mom}
\boldsymbol{k}=\boldsymbol{e_{\langle t \rangle}}+\boldsymbol{e_{\langle r \rangle}} \,.
\end{equation}
In the Kerr geometry (\ref{kerr_metric}), the timelike unit surface four-velocity, $\boldsymbol{U_{\mathrm{surf}}}$, of the stationary observer is
\begin{equation}
\boldsymbol{U_{\mathrm{surf}}}=\mathcal{N}\,(\boldsymbol{\partial_t}+\Omega_{\star}\,\boldsymbol{\partial_\varphi})\,, 
\end{equation}
where
\begin{equation}
\mathcal{N}=\frac{1}{\sqrt{-\mathrm{g_{tt}}-\Omega_{\star}  (2\, \mathrm{g_{t\varphi}}+\mathrm{g_{\varphi \varphi}} \Omega_{\star})}}\,.
\end{equation}
Here the metric coefficients are evaluated at the radial coordinate $\RS$ and at the polar coordinate of the photon emission $\theta_e$. The corotating orthonormal frame components related to the stationary observer sitting at the radiating surface take the following form 
\begin{equation} \label{surfaceframe}
\begin{aligned}
\boldsymbol{e_{\langle t \rangle}}&=\boldsymbol{U_{\mathrm{surf}}},\
\boldsymbol{e_{\langle r \rangle}}=\frac1{\sqrt{g_{rr}}}\boldsymbol{\partial_r},\
\boldsymbol{e_{\langle \theta \rangle}}=\frac1{\sqrt{g_{\theta \theta }}}\boldsymbol{\partial_\theta},\\
\boldsymbol{e_{\langle \phi \rangle}}&=\frac{\mathcal{N}\left[\left(\mathrm{g_{t\varphi}}+\mathrm{g_{\varphi \varphi}} \Omega_{\star}\right) \boldsymbol{\partial_t}+\left(\mathrm{g_{tt}}+\mathrm{g_{t \varphi}} \Omega_{\star}\right) \boldsymbol{\partial_\varphi} \right]}{\sqrt{\mathrm{g_{t\varphi}}^2-\mathrm{g_{\varphi\varphi}} \mathrm{g_{tt}}}}.
\end{aligned}
\end{equation}
Then, the photon four-momentum (\ref{local_mom}) can be rewritten in such coordinate frame as 
\begin{equation} \label{coordinate_mom}
\boldsymbol{k}=\mathcal{N}\boldsymbol{\partial_t}+ \frac1{\sqrt{g_{rr}}}\boldsymbol{\partial_r}+ \Omega_{\star}\mathcal{N}\boldsymbol{\partial_\varphi}\,. 
\end{equation}
A straightforward calculation yields the azimuthal impact parameter $b=-k_{\varphi}/k_{t}$ 
\begin{eqnarray} 
&&b\equiv-\frac{\mathrm{g_{t\varphi}}+\mathrm{g_{\varphi\varphi}}\Omega_{\star} }{\mathrm{g_{tt}}+\mathrm{g_{t\varphi}} \Omega_{\star}}=\label{kerr_impact_parameter}\\  
&&= \frac{\sin ^2\theta_e \left[\left(\RS^4+a^2\mathcal{B}\right)\Omega_{\star}-2Ma\RS\right]}
       {\left[\RS^2-2M \RS+a\left(a\cos ^2\theta_e+ 2M\RS\Omega_{\star} \sin ^2\theta_e\right)\right]},\notag
\end{eqnarray}
where
\begin{equation}
\mathcal{B}=\left(a^2+\RS^2\right) \cos^2\theta_e+2M \RS \sin^2\theta_e+\RS^2
\end{equation}
$b$ can be approximated in terms of the photon impact parameter in the Schwarzschild geometry, $b_{\rm{Schw}}$, plus a correction term of first order in the spin parameter  
\begin{equation} \label{apprx_b}
b=b_{\rm{Schw}}-\mathcal{A}\,a+O\left(a^2\right), 
\end{equation}
where 
\begin{eqnarray} \label{schw_impact_parameter}
&&b_{\mathrm{Schw}}=\frac{\RS^2\Omega_{\star} \sin^2\theta_e}{1-2M/\RS}, \\
&&\mathcal{A}= \frac{2M\sin^2\theta_e\left(M\RS^3 \Omega_{\star}^2 \sin^2\theta_e+\RS-2M \right)}{(\RS-2M)^2}.  \label{first_der}
\end{eqnarray} 
From Eq. (\ref{apprx_b}) it is apparent that the radially outgoing photon has the same specific angular momentum of a matter element located at the emission point on the rotating surface (see Eq. (\ref{schw_impact_parameter})). In addition, the influence of frame-dragging entails the decrease of the value of the emitted photon impact parameter of the radiation field (see Eq. (\ref{first_der})). The azimuthal photon impact parameter is not fixed, but instead it ranges from the minimum value $b(\theta_e=0)=0$ (describing photons emitted from the poles of the surface along polar axis) to the maximum value $b(\theta_e=\pi/2)$ (corresponding to photons emitted in the equatorial plane). 

The latitudinal photon impact parameter, $q$, is obtained from the condition 
\begin{equation} \label{cond2}
\Theta_{b,q}(\theta_e)=0,
\end{equation}
resulting from the absence of the polar component of the photon four-momentum (\ref{coordinate_mom}). From Eqs. (\ref{Thetapot}) and (\ref{cond2}) we can express  $q$ as a function of the photon emission polar angle $\theta_e$ as
\begin{equation} \label{q_r}
q=b^{2}\cot^{2} \theta_e-a^{2} \cos^{2}\theta_e\ . 
\end{equation}
This relation also fixes the Carter constant $\mathcal{Q}$ for a given photon trajectory through Eq. (\ref{Carter_constant}).
The latitudinal potential (\ref{Thetapot}) is independent of the radial coordinate and therefore the polar angle, $\theta=\theta_e$, along a given photon trajectory is conserved. 
Therefore in the case of our radiation field which is emitted radially in the frame of the rigidly rotating emitting surface, the latitudinal photon impact parameter $q$ (and also the Carter constant $\mathcal{Q}$) is fully determined by the emission angle $\theta_e$, the azimuthal photon impact parameter $b(\theta_e,\Omega_{\star},\RS, a)$ and the Kerr parameter (spin) $a$. As in our previous paper \cite{DeFalco20183D}, such setup for the radiation field significantly simplifies the integration of test particle trajectories in that only one photon trajectory (described by the corresponding pair $(b,q)$) emitted at $\theta_e$ can reach the instantaneous test particle position at the polar coordinate $\theta = \theta_e$. 

Naturally, also in the arbitrary ZAMO frame (\ref{eq:zamoframes}), the local polar component of the photon four-momentum remains identically zero ($k^{\hat \theta}=k^{\theta}/\sqrt{g_{\theta\theta}} = 0 $). Then using Eq. (\ref{photon}), we can easily see that the local polar angle of the photon four-momentum in the ZAMO frame always takes the value $\zeta=\pi/2$. In view of the last condition, we have that the azimuthal photon angle $\beta$, given by Eq. (\ref{ANG1}), takes the form
\begin{equation} \label{ANG12}
\cos\beta=\frac{b N}{\sqrt{g_{\varphi\varphi}}(1+b N^\varphi)}.
\end{equation}

We adopt the radiation field we introduced in this section as a heuristic approximation of  
the emission from a rotating NS  or from the surface of a rotating hot corona in the close vicinity 
of a BH, and therefore we 
will consider hereafter only the case of outgoing photons, {\it i.e.} $\sin\beta \geq 0$.
Considering only the emission of an individual light ray (that orthogonal to the surface) 
from each point of the rotating sphere, as done in the present study,  
makes the problem analytically treatable and represents a clear step forward in capturing the behaviour 
of test particle under the influence of the gravitational and radiation field generated by (or around) 
rotating compact objects. More realistic models should include the 
the whole range of outgoing light ray directions as well as the angular dependence of the surface emissivity.

\subsubsection{Emission from NS surface}
\label{sec:NS_surface}
Let us consider a radiation field emitted from the surface a rotating NS. The Kerr parameter $a$ 
is expressed as a function the NS angular velocity $\Omega_{\star}$ through 
\begin{equation} \label{eq:NSa} 
a=\mathcal{C}\Omega_{\star}\,, 
\end{equation}
where $\mathcal{C}$ depends on the NS structure and equation of state \cite{Haensel2009,Urbanec2010,Rikovska2003}. For a NS of radius $\RS=11\,\mathrm{km}$ and mass $M=1.5\,\mathrm{M_{\odot}}$, we choose $\mathcal{C}=1.1\,.\,10^{-4}\,\mathrm{s}$ (see, e.g., \cite{Haensel2009,Urbanec2010,Rikovska2003,Bakala2012} for details). A moderate NS rotation frequency of $f_{\star}\equiv\Omega_{\star}/2\pi=300\,\mathrm{Hz}$ corresponds to $a \approx 0.21$. In this case the maximum value of the azimuthal photon impact parameter on the NS equator would be $b(\theta_e=\pi/2) \approx 0.43$. For a NS rotation frequency of $f_{\star} = 900\,\mathrm{Hz}$  we get $a \approx 0.62$ and  $b(\theta_e=\pi/2) \approx 1.29$. 
In the non-rotating case ($a=0$, Schwarzschild spacetime) the radiation field has $b=0$, which is identical with the zero spin test field described in \cite{DeFalco20183D}. Note, that the physically meaningful range of rotation frequency is limited by the value of the NS break-uo frequency from above.

\subsubsection{Emission from a hot corona around a BH} 
\label{sec:BH_corona}
Let us now consider a radiation field emitted from a spherical corona which is rigidly rotating with angular velocity $\Omega_{\mathrm{\star}}$. We assume that the Kerr parameter, $a$, is determined entirely by the central BH. We restrict our consideration here only positive values of the angular velocity $\Omega_{\mathrm{\star}}$, {\it i.e.} emitting surfaces corotating with the BH.

The existence of the emission surface rotating with $\Omega_{\star}\ge0$ and located on $\RS$ requires that the four-velocity $\boldsymbol{U_{\mathrm{surf}}}$ takes real values. This corresponds to positive values of $\mathcal{N}$ on the equatorial plane ($\theta_e=\pi/2$), where $\boldsymbol{U_{\mathrm{surf}}}$ reaches its maximum value. Such condition translates into
\begin{equation}
\label{superluminal}
\begin{aligned}
&\Omega_{-}(\RS,\pi/2) < \Omega_{\star} < \Omega_{+}(\RS,\pi/2)\ ,\\
&\Omega_{\pm}(\RS,\pi/2)=\frac{-\mathrm{g_{t\varphi}} \pm \sqrt{ \mathrm{g_{t\varphi}}^2 - \mathrm{g_{\varphi\varphi}} \mathrm{g_{tt}}}}{\mathrm{g_{\varphi\varphi}}}\ ,
\end{aligned}
\end{equation}
which exclude the possibility of superluminal rotation of the emission surface. Note that in the investigated case of $\Omega_{\star} >0 $, the minimum value $\Omega_{-}(\RS,0)$ is relevant only inside of the BH ergosphere, where there cannot be static observers. 

\begin{figure*}[htbp]  
\begin{center}
\includegraphics[width=1.00\linewidth, angle=0]{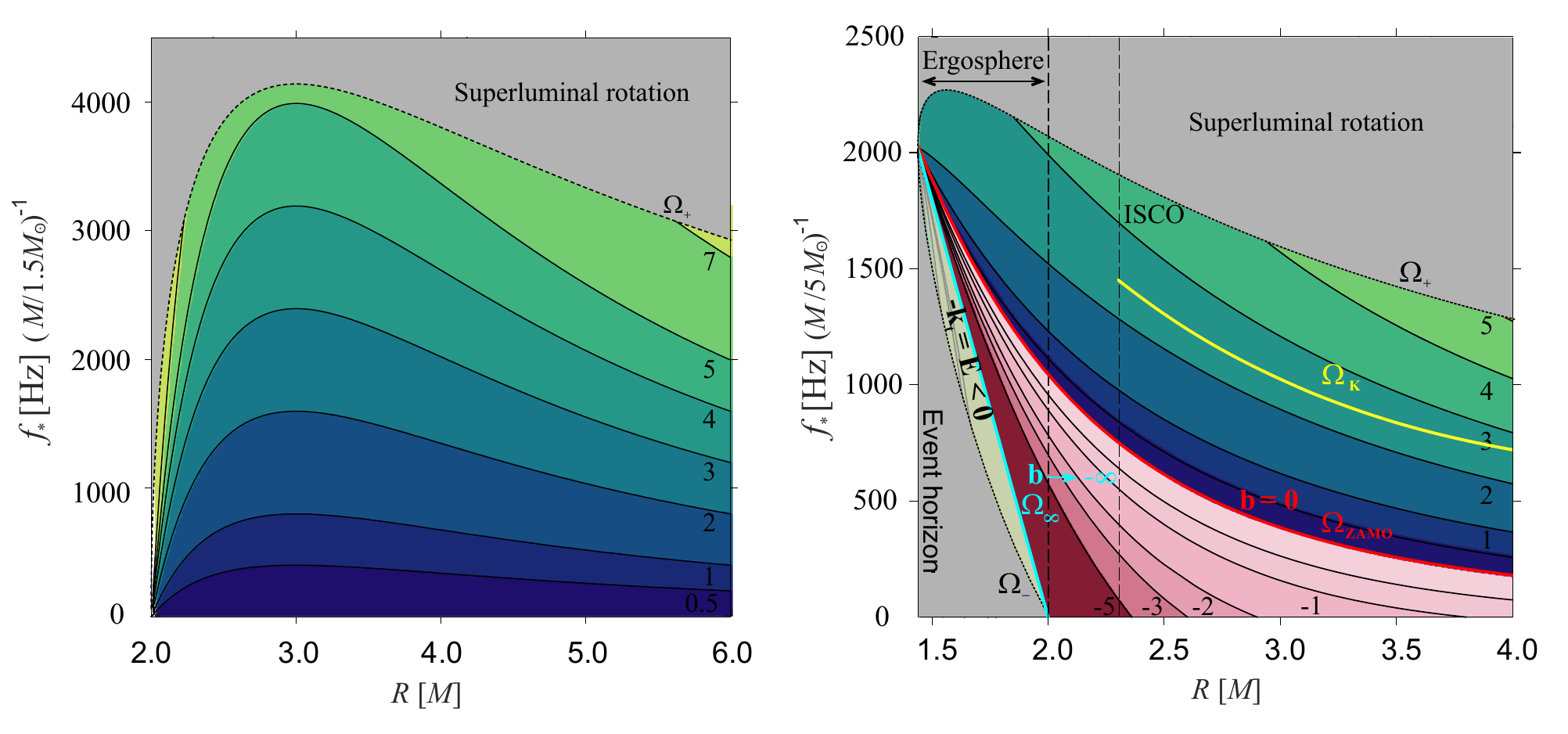}
\end{center}
\caption{Contour plot of the azimuthal impact parameter $b$ of photons emitted from the equator ($\theta_e=\pi/2$) as  function of radius, $\RS$, and rotation frequency, $f_{\star}=\Omega_{\star}/2\pi$ of the emitting surface. Left panel: Schwarzschild geometry using a NS mass of $M=1.5\mathrm{M_\odot}$. Positive impact parameter values are limited from above only by $\Omega_{+}(\RS,\pi/2)$. Right panel: Kerr geometry using a value of $a=0.9$ and a BH mass of $M=5\mathrm{M_\odot}$. The green region corresponds to positive values of $b$, while the red region corresponds to negative values. The yellow curve ending at the radius of the innermost stable cicular orbit (ISCO) denotes the Keplerian orbital frequency $\Omega_{K}=[M(r^{3/2}+a)]^{-1}$. The red curve denotes the zero value of $b$ corresponding to the condition $\Omega_{\star}=\Omega_{\mathrm{ZAMO}}(\RS)$, {\it i.e.} the surface rotates at the frequency of spacetime rotation. Along the cyan curve, where $\Omega_{\star}=\Omega_{\infty}(\RS)\equiv-\mathrm{g_{tt}}/\mathrm{g_{t\varphi}}$, the value of the azimuthal photon impact parameter diverges ($b \to -\infty$). The beige region under the cyan curve corresponds to photons emitted with negative conserved energy $-k_t$ from surfaces rotating at $\Omega_{-}(\RS,0) < \Omega_{\star} \le \Omega_{\infty}(\RS)$. Such photons cannot escape from the ergosphere. The gray regions denote the cases of (unphysical) superluminal rotation.}  
\label{fig:lambdarange}
\end{figure*}

\subsubsection{$b$-range in the Schwarzschild geometry}
\label{sec:brange_schw}
In the Schwarzschild case, the photon impact parameter values (\ref{schw_impact_parameter}) are always positive and limited from above only by the condition $\Omega_{\star} < \Omega_{+}(\RS,\pi/2)$ (see left panel in Fig. \ref{fig:lambdarange}). The maximum frequency $\Omega_{+}(\RS,\pi/2)$ is indeed unphysical. If we consider a emitting rotating hot corona in the background of the Schwarzchild geometry, the rotation frequency $f_{\star}=300\,\mathrm{Hz}$ corresponds to the value of $b_{\mathrm{Schw}} \approx 0.575$, while the higher rotation frequency $f_{\star} = 900\,\mathrm{Hz}$ corresponds to the value of $b_{\mathrm{Schw}} \approx 1.725$. 

\subsubsection{Estimation of $b$-range in Kerr geometry}
\label{sec:brange_kerr}

In the Kerr metric, frame-dragging makes matters more intricate. The upper limit of $b$ is again controlled by the condition $\Omega_{\star} < \Omega_{+}(\RS,\pi/2)$, but contrary to the Schwarzschild case, the azimuthal photon impact parameter $b$ can be both positive and negative. We can distinguish the following cases for the behaviour of the azimuthal photon impact parameter $b$ (see the right panel in Fig. \ref{fig:lambdarange}):

\begin{itemize}
\item for $\Omega_{\star}=\Omega_{\mathrm{ZAMO}}(\RS)$ $b=0$, as the emitting surface rotates with the same angular velocity as the spacetime (red curve); 
\item for $\Omega_{\star} \ge \Omega_{\mathrm{ZAMO}}(\RS)$, $b\ge0$ (green region); 
\item for $\Omega_{\star} \le \Omega_{\mathrm{ZAMO}}(\RS)$ $b\le0$ (red region);
\end{itemize}
If the emitting surface is located inside the ergosphere, where the surface rotation frequency is limited by the condition $0 < \Omega_{-}(\RS,0) <  \Omega_{\star}$, we can distinguish the following additional cases:  
\begin{itemize}
\item  for $\Omega_{\infty}(\RS) \le \Omega_{\star} \le \Omega_{\mathrm{ZAMO}}(\RS)$ photons can possess extreme negative values of $b$ reaching $-\infty$ for $\Omega_{\star} \rightarrow \Omega_{\infty}(\RS)$ (see the bottom part and boundary of the red region).
\item for $\Omega_{\star}=\Omega_{\infty}(\RS)$ with $\Omega_{\infty}(\RS)\equiv-\mathrm{g_{tt}}/\mathrm{g_{t\phi}}>\Omega_{-}(\RS,0)$, the conserved photon energy is $k_t=0$ and $\lim_{\Omega_{\star}\rightarrow\Omega_{\infty}^\pm}b(\theta_e=\pi/2)=\mp \infty$ (see the cyan curve);
\item for $\Omega_{-}(\RS,0) < \Omega_{\star} \le \Omega_{\infty}(\RS)$, the conserved photon energy is negative $k_t<0$, and photons cannot escape from the ergosphere. Their $b$ is positive, reaching $+\infty$ for $\Omega_{\star}\rightarrow \Omega_{\infty}(\RS)$. Such radiation fields cannot affect physical processes outside the ergosphere (see the beige region and \cite{Grib2017} for further details). 
\end{itemize}

\subsection{Intensity parameter}
\label{sec:int_par}
Since the photon four-momentum $\boldsymbol{k}$ is completely determined by the $(b,q)$ pair, the coordinate dependence of $\Phi$ then follows from the conservation equations $T^{\alpha\beta}{}_{;\beta}=0$. Due to the absence of photon latitudinal motion ($k^{\theta}=0$) and the symmetries of the Kerr spacetime, these can be written as
\begin{equation}\label{flux_cons}
0=\nabla_\beta (\Phi^2 k^\beta)=\frac{1}{\Sigma\sin\theta}\,\partial_r (\Sigma\sin\theta\,\Phi^2 k^r)\,. 
\end{equation}

However, Eq. (\ref{flux_cons}) does not fix the intensity parameter unambiguously.
The conservation equations would still be fulfilled, even if we multiplied the  
derivative with respect to the radial coordinate ($\Sigma\sin\theta\,\Phi^2 k^r $) of the expression by an arbitrary function of the $\theta$ coordinate. Therefore condition (\ref{flux_cons}) determines the class of radiation fields that differ from one another by the latitudinal distribution of intensity. A natural requirement for the radiation field model is that it attains spherical symmetry for the case of pure radially outgoing field ($b=0$) in a Schwarzschild background geometry. This in turn requires that  the intensity parameter is independent of the $\theta$ coordinate, which can be easily achieved by multiplying the derivative with respect to the radial coordinate of (\ref{flux_cons}) by a factor $1/\sin\theta$.
The condition (\ref{flux_cons}) then will take the form
\begin{equation}
0=\partial_r (\Sigma\Phi^2 k^r)\,. 
\end{equation}
By expressing $k^r$ through Eq. (\ref{CarterEQs}) and substituting the values of the impact parameters obtained above in Eqs. (\ref{kerr_impact_parameter}) and (\ref{q_r}) in Eq. (\ref{flux_cons}), we obtain
\begin{equation} \label{eq:phi}
\Phi^2 \sqrt{R_{b,q}(r)}=\hbox{\rm const} = \Phi_0^2\,,
\end{equation}
where $\Phi_0^2$ is a new constant related to the intensity of the radiation field at the emitting surface.
Then, the parameter $\Phi^2$ is related to the radial effective potential (\ref{Rpot}) by the formula
\begin{equation}\label{INT_PAR}
\Phi^2=\frac{\Phi_0^2}{\sqrt{R_{b,q}(r)}}\,.
\end{equation}
Note that in the case of a non-rotating emitting surface in a Schwarzschild spacetime, Eq. (\ref{INT_PAR}) has the simple limit $\Phi^2=\Phi_0^2/r^2$,
apparently preserving the spacetime spherical symmetry. Also note, that in the case $b=0$ expression (\ref{INT_PAR}) coincides with the formula for the intensity parameter of the radial radiation field described in \cite{DeFalco20183D} and can be considered its generalization.

\section{Interaction of radiation - test particle }
\label{sec:eoms}

\subsection{Test particle motion}
\label{sec:test_part}
We consider a test particle moving in the 3D space endowed with a timelike four-velocity $\boldsymbol{U}$ and a spatial three-velocity with respect to the ZAMOs, $\boldsymbol{\nu}(U,n)$, \cite{DeFalco20183D}:
\begin{eqnarray} 
&&\boldsymbol{U}=\gamma(U,n)[\boldsymbol{n}+\boldsymbol{\nu}(U,n)], \label{testp}\\
&&\boldsymbol{\nu}=\nu(\sin\psi\sin\alpha\boldsymbol{e_{\hat r}}+\cos\psi\boldsymbol{e_{\hat\theta}}+\sin\psi \cos\alpha \boldsymbol{e_{\hat\varphi}}),
\end{eqnarray}
where $\gamma(U,n)=1/\sqrt{1-||\boldsymbol{\nu}(U,n)||^2}$ is the Lorentz factor, $\nu=||\boldsymbol{\nu}(U,n)||$, $\gamma(U,n) =\gamma$. Here $\nu$ represents the magnitude of the test particle spatial velocity $\boldsymbol{\nu}(U,n)$, $\alpha$ is the azimuthal angle of the vector $\boldsymbol{\nu}(U,n)$ measured clockwise from the positive $\hat\varphi$ direction in the $\hat{r}-\hat{\varphi}$ tangent plane in the ZAMO frame, and $\psi$ is the polar angle of the vector $\boldsymbol{\nu}(U,n)$ measured from the axis orthogonal to the $\hat{r}-\hat{\varphi}$ tangent plane in the ZAMO frame (see Fig. \ref{fig:Fig0}). The explicit expression for the test particle velocity components with respect to the ZAMOs are 
\begin{equation} 
\begin{aligned}\label{four_velocity}
&U^t\equiv \frac{dt}{d\tau}=\frac{\gamma}{N},\quad U^r\equiv \frac{dr}{d\tau}=\frac{\gamma\nu^{\hat r}}{\sqrt{g_{rr}}},\\
&U^\theta\equiv \frac{d\theta}{d\tau}=\frac{\gamma\nu^{\hat\theta}}{\sqrt{g_{\theta\theta}}},\quad U^\varphi\equiv \frac{d\varphi}{d\tau}=\frac{\gamma\nu^{\hat\varphi}}{\sqrt{g_{\varphi\varphi}}}-\frac{\gamma N^\varphi}{N} ,
\end{aligned}
\end{equation}
where $\tau$ is the proper time parameter along $\bold{U}$. Using the \emph{observer-splitting formalism}, the test particle acceleration relative to the ZAMO congruence, $\boldsymbol{a}(U)=\nabla_{\bold U} \bold{U}$, is given by \cite{Defalco2018,DeFalco20183D}
\begin{eqnarray}
a(U)^{\hat r}&=& \gamma^2 [a(n)^{\hat r}+k_{\rm (Lie)}(n)^{\hat r}\,\nu^2 (\cos^2\alpha\sin^2\psi\label{acc1} \\
&&+\cos^2\psi)+2\nu\cos \alpha\sin\psi\, \theta(n)^{\hat r}{}_{\hat \varphi}]\nonumber\\
&&+\gamma \left(\gamma^2 \sin\alpha\sin\psi \frac{\rm d \nu}{\rm d \tau}+\nu \cos \alpha\sin\psi \frac{\rm d \alpha}{\rm d \tau}\right.\nonumber\\
&&\left.+\nu \cos \psi\sin\alpha \frac{\rm d \psi}{\rm d \tau} \right), \nonumber\\   
a(U)^{\hat \theta}&=&\gamma^2 [a(n)^{\hat \theta}+k_{\rm (Lie)}(n)^{\hat \theta}\,\nu^2 \sin^2\psi\cos^2\alpha\label{acc3}\\
&&-k_{\rm (Lie)}(n)^{\hat r}\, \nu^2\sin\psi\sin\alpha\cos\psi\nonumber\\
&&+2\nu\cos \alpha\sin\psi\, \theta(n)^{\hat \theta}{}_{\hat \varphi}]\nonumber\\
&&+ \gamma\left(\gamma^2 \cos\psi \frac{\rm d \nu}{\rm d \tau}-\nu \sin\psi \frac{\rm d \psi}{\rm d \tau}\right).\nonumber\\
a(U)^{\hat \varphi}&=& -\gamma^2 \nu^2\cos \alpha\sin\psi\left[ \sin \alpha \sin\psi\, k_{\rm (Lie)}(n)^{\hat r}\right.\label{acc2}\\ 
&&\left.+ k_{\rm (Lie)}(n)^{\hat \theta}\cos\psi\right]+ \gamma\left(\gamma^2 \cos \alpha\sin\psi \frac{\rm d \nu}{\rm d \tau}\right.\nonumber\\
&&\left.-\nu\sin \alpha\sin\psi \frac{\rm d \alpha}{\rm d \tau}+\nu\cos\alpha\cos\psi \frac{\rm d \psi}{\rm d \tau}\right),\nonumber\\
&&a(U)^{\hat t}=\gamma^2\nu\left\{\sin \alpha \sin\psi\left[a(n)^{\hat r}\right.\right.\label{acc4}\\
&&\left.\left.+2\nu\cos \alpha\sin\psi\, \theta(n)^{\hat r}{}_{\hat \varphi}\right] \right. \nonumber\\
&&\left.+\cos\psi\left[a(n)^{\hat \theta}\right.\right.\nonumber\\
&&\left.\left.+2\nu\cos\alpha\sin\psi \theta(n)^{\hat \theta}{}_{\hat \varphi}\right]\right\}+ \gamma^3\nu \frac{\rm d \nu}{\rm d \tau}\nonumber. 
\end{eqnarray}

\subsection{Splitting of the radiation force}
\label{sec:split_rad_field}
We assume that the radiation test particle interaction occurs through elastic, Thomson-like scattering, characterized by a constant momentum-transfer cross section $\sigma$, independent of direction and frequency of the radiation field. The radiation force is \cite{DeFalco20183D}
\begin{equation} \label{radforce}
{\mathcal F}_{\rm (rad)}(U)^\alpha = -\sigma P(U)^\alpha{}_\beta \, T^{\beta}{}_\mu \, U^\mu \,,
\end{equation}
where $P(U)^\alpha{}_\beta=\delta^\alpha_\beta+U^\alpha U_\beta$ projects a vector orthogonally to $\bold{U}$. Decomposing the photon four-momentum $\bold{k}$ first with respect to the test particle four-velocity, $\bold{U}$, and then in the ZAMO frame, $\bold{n}$, we have \cite{DeFalco20183D}
\begin{equation} \label{diff_obg}
\bold{k} = E(n)[\bold{n}+\boldsymbol{\hat{\nu}}(k,n)]=E(U)[\bold{U}+\boldsymbol{\hat {\mathcal V}}(k,U)].
\end{equation}
Using Eq. (\ref{diff_obg}) in Eq. (\ref{radforce}), we obtain
\begin{equation} \label{Frad0}
\begin{aligned}
{\mathcal F}_{\rm (rad)}(U)^\alpha&=-\sigma \Phi^2 [P(U)^\alpha{}_\beta k^\beta]\, (k_\mu U^\mu)\\
&=\sigma \, [\Phi E(U)]^2\, \hat {\mathcal V}(k,U)^\alpha\,.
\end{aligned}
\end{equation}
The test particle equations of motion is $m \bold{a}(U) = \boldsymbol{{\mathcal F}_{\rm (rad)}}(U)$, where $m$ is the test particle mass;  by defining $\tilde \sigma=\sigma/m$, we have 
\begin{equation}\label{geom}
\bold{a}(U)=\tilde \sigma \Phi^2 E(U)^2  \,\boldsymbol{\hat {\mathcal V}}(k,U).
\end{equation} 
Taking the scalar product of Eq. (\ref{diff_obg}) and 
 $\bold{U}$ and considering $\zeta=\pi/2$, we find 
\begin{equation} \label{enepart}
E(U)=\gamma E(n)[1-\nu\sin\psi\cos(\alpha-\beta)],
\end{equation}
Such splitting permits to determine $\boldsymbol{\hat{\mathcal{V}}}(k,U)=\hat{\mathcal{V}}^t\boldsymbol{n}+\hat{\mathcal{V}}^r\boldsymbol{e_{\hat r}}+\hat{\mathcal{V}}^\theta \boldsymbol{e_{\hat\theta}}+\hat{\mathcal{V}}^\varphi \boldsymbol{e_{\hat\varphi}}$ as \cite{DeFalco20183D}
\begin{eqnarray}
&&\hat{\mathcal{V}}^{\hat r}=\frac{\sin\beta}{\gamma [1-\nu\sin\psi\cos(\alpha-\beta)]}-\gamma\nu\sin\psi\sin\alpha,\label{rad1}\\
&&\hat{\mathcal{V}}^{\hat \theta}=-\gamma\nu\cos\psi \label{rad2},\\
&&\hat{\mathcal{V}}^{\hat\varphi}=\frac{\cos\beta}{\gamma [1-\nu\sin\psi\cos(\alpha-\beta)]}-\gamma\nu\sin\psi\cos\alpha,\label{rad3}\\
&&\hat{\mathcal{V}}^{\hat t}=\gamma\nu\left[\frac{\sin\psi\cos(\alpha-\beta)-\nu}{1-\nu\sin\psi\cos(\alpha-\beta)}\right].\label{rad4}
\end{eqnarray}

\subsection{General relativistic equations of motion}
The equations of motion for a test particle moving in a 3D space, defined in terms of $(r,\theta,\varphi,\nu,\psi,\alpha)$, represent a set of coupled ordinary differential equations of the first order, i.e., \cite{DeFalco20183D}
\begin{eqnarray}
&&\frac{d\nu}{d\tau}= -\frac{1}{\gamma}\left\{ \sin\alpha \sin\psi\left[a(n)^{\hat r}\right.\right.\label{EoM1}\\
&&\left.\left.\ \quad +2\nu\cos \alpha\sin\psi\, \theta(n)^{\hat r}{}_{\hat \varphi} \right]+\cos\psi\left[a(n)^{\hat \theta}\right. \right.\nonumber\\
&&\left.\left.\ \quad+2\nu\cos\alpha\sin\psi\, \theta(n)^{\hat \theta}{}_{\hat \varphi}\right]\right\}+\frac{\tilde{\sigma}[\Phi E(U)]^2}{\gamma^3\nu}\hat{\mathcal{V}}^{\hat t},\nonumber\\
&&\frac{d\psi}{d\tau}= \frac{\gamma}{\nu} \left\{\sin\psi\left[a(n)^{\hat \theta}+k_{\rm (Lie)}(n)^{\hat \theta}\,\nu^2 \cos^2\alpha\right.\right.\label{EoM2}\\
&&\left.\left.\ \quad+2\nu\cos \alpha\sin^2\psi\ \theta(n)^{\hat \theta}{}_{\hat \varphi}\right]-\sin \alpha\cos\psi \left[a(n)^{\hat r}\right.\right.\nonumber\\
&&\left.\left.\ \quad+k_{\rm (Lie)}(n)^{\hat r}\,\nu^2+2\nu\cos \alpha\sin\psi\, \theta(n)^{\hat r}{}_{\hat \varphi}\right]\right\}\nonumber\\
&&\ \quad+\frac{\tilde{\sigma}[\Phi E(U)]^2}{\gamma\nu^2\sin\psi}\left[\hat{\mathcal{V}}^{\hat t}\cos\psi-\hat{\mathcal{V}}^{\hat \theta}\nu\right],\nonumber\\
&&\frac{d\alpha}{d\tau}=-\frac{\gamma\cos\alpha}{\nu\sin\psi}\left[a(n)^{\hat r}+2\theta(n)^{\hat r}{}_{\hat \varphi}\ \nu\cos\alpha\sin\psi\right.\label{EoM3}\\
&&\left.\ \quad+k_{\rm (Lie)}(n)^{\hat r}\,\nu^2+k_{\rm (Lie)}(n)^{\hat \theta}\,\nu^2\cos^2\psi \sin\alpha\right]\nonumber\\
&&\ \quad+\frac{\tilde{\sigma}[\Phi E(U)]^2\cos\alpha}{\gamma\nu\sin\psi}\left[\hat{\mathcal{V}}^{\hat r}-\hat{\mathcal{V}}^{\hat \varphi}\tan\alpha\right],\nonumber\\
&&U^r\equiv\frac{dr}{d\tau}=\frac{\gamma\nu\sin\alpha\sin\psi}{\sqrt{g_{rr}}}, \label{EoM4}\\
&&U^\theta\equiv\frac{d\theta}{d\tau}=\frac{\gamma\nu\cos\psi}{\sqrt{g_{\theta\theta}}} \label{EoM5},\\
&&U^\varphi\equiv\frac{d\varphi}{d\tau}=\frac{\gamma\nu\cos\alpha\sin\psi}{\sqrt{g_{\varphi\varphi}}}-\frac{\gamma N^\varphi}{N},\label{EoM6}
\end{eqnarray}
For $b=0$ the equations of motion reduce to the radial radiation field in 3D case \cite{DeFalco20183D}.

We define the relative luminosity of the radiation field as $A/M=\tilde{\sigma}\Phi_0^2E^2\equiv L_\infty/L_{\rm Edd}$, ranging in $[0,1]$, where $L_\infty$ is the luminosity at infinity and $L_{\rm EDD}=4\pi Mm/\sigma$ is the is the Eddington luminosity at infinity \cite{Bini2009,Bini2011,DeFalco20183D}. Using Eqs. (\ref{INT_PAR}) and (\ref{enepart}), the term $\tilde{\sigma}[\Phi E(U)]^2$ becomes
\begin{equation} \label{eq: sigma_tilde}
\tilde{\sigma}[\Phi E(U)]^2=\frac{ A\,\gamma^2(1+b N^\varphi)^2[1-\nu\sin\psi\cos(\alpha-\beta)]^2}{N^2\sqrt{R_{b,q}(r)}}.
\end{equation}

In Appendix \ref{Appendix_Classic_Limit} we show the way Eqs. (\ref{EoM1})--(\ref{EoM6}) reduce in the classical limit. In Appendix \ref{Appendix_Weak_Field} we show the weak field approximations of Eqs. (\ref{EoM1})--(\ref{EoM6}) at the first order in the Kerr parameter $a$ for slow rotations ($a\to0$).

\section{Critical hypersurfaces}
\label{sec:critc_rad}
The set of Eqs. (\ref{EoM1})--(\ref{EoM6}) admits a solution in which gravitational attraction, radiation pressure, and PR drag effect balance each other on an axially symmetric hypersurface, partially or fully encapsulating the emitting sphere. In order to obtain the condition determining the radial coordinate of such a hypersurface, the critical radius $r_{\rm (crit)}$, we consider a captured test particle in radial equilibrium moving in a purely circular motion ($\alpha=0\,;\psi=\pm\pi/2;\nu=\mbox{const}$) along,  in general, a non-equatorial orbit. Then for $\frac{d\nu}{d\tau}=0$ Eq. (\ref{EoM1}) takes the following form
\begin{equation}\label{eq:crit_hyper1}
\frac{A(1+bN^\varphi)^2(1 - \nu\cos\beta)(\cos\beta-\nu)}{N^2 \sqrt{R_{b,q}(r_{\rm (crit))}}}=0,
\end{equation}
from which it follows 
\begin{equation}\label{eq:crit_veloc}
\nu=\cos\beta.
\end{equation}
Therefore the azimuthal velocity of the captured test particle is equal to the azimuthal velocity of the photons of the radiation field\footnote{Assuming the opposite direction of the particle velocity ($\alpha=\pi$), we get at the same solution except for the negative sign of $\nu$.}.
Considering that in the ZAMO frame the velocity of the captured test particle is tangential to the hypersurface ($\alpha=0$), for $\frac{d\alpha}{d\tau}=0$ Eq. (\ref{EoM3}) takes the following form
\begin{equation}\label{eq:crit_hyper2}
\begin{aligned}
&a(n)^{\hat r} + 2\theta(n)^{\hat r}{}_{\hat\varphi}\nu+k_{\rm (Lie)}(n)^{\rm r}\nu^2 =\\
&=\frac{A(1+bN^\varphi)^2\sin\beta}{N^2\gamma^3 \sqrt{R_{b,q}(r_{\rm (crit))}}}.
\end{aligned}
\end{equation}
This generalises the result obtained in \cite{Bini2011}, where the equilibrium was analyzed only in the equatorial plane. Note that, similarly to the 3D case with purely radial radiation field \cite{DeFalco20183D}, a latitudinal drift  on the critical hypersuface towards the equatorial plane is to be expected, as the condition $\frac{d\psi}{d\tau}=0$ is not fulfilled automatically. In Sec. \ref{sec:EMCH} we analyse in detail the condition for the existence of orbits with a constant latitude coordinate. Note also that in the case of a radiation field with zero angular momentum ($b=0\,;\beta=\pi/2$), Eq. (\ref{eq:crit_hyper2}) reduces to Eq. (53) in \cite{DeFalco20183D}; for non-zero angular momentum radiation fields 
the same holds on the polar axis ($\theta=0,\pi$) where the angular momentum of the photons emitted by the spherical surface is zero. Therefore, the axisymmetric shape of the critical hypersurfaces can be obtained by solving the implicit equation (\ref{eq:crit_hyper2}) for the $(r,\theta)$ coordinates by specifying a set of the initial parameters $\left\{A,a,R_\star,\Omega_\star\right\}$. The pair of photon impact parameters $(b,q)$ is found in terms of $R_\star,\Omega_\star,\theta$  through Eq. (\ref{kerr_impact_parameter}) and Eq. (\ref{q_r}), respectively \footnote{Equation (\ref{q_r}) displays a divergence in $q$  at the polar axis (i.e., $\theta=0,\pi$). Therefore our numerical analysis of the behavior extends to the close vicinity of, rather than the poles themselves.}.

In our previous work \cite{DeFalco20183D}, in which a zero angular momentum radiation field ($b=0\,;\beta=\pi/2$) was analyzed, we found that for non-zero $a$ the critical hypersurface always takes a prolate shape due to the fact the frame dragging effect attains its maximum at the equator. In the more general case considered in the present study, the shape of the critical surface, besides frame dragging, is determined also by the angular momentum of the radiation field photons, which is now a function of the polar coordinate. Hereafter we show that the critical hypersurfaces may morph between an oblate and a prolate shape depending on the parameters governing the value of the critical radius $r_{\rm (crit)}=r_{\rm (crit)}(A,a,\theta,\Omega_\star,R_\star)$. 
We carry out a more detailed analysis of the properties of the critical hypersurface in the  cases of a NS and a BH in Secs. \ref{sec:CHNS} and \ref{sec:CHBH}, respectively.

\subsection{Multiplicity of the critical hypersurface}
\label{sec:EMCH}

As demonstrated in \cite{Bini2011}, Eq. (\ref{eq:crit_hyper2}) may have one or three solutions depending on the values of $A$,$a$,$b$. The switch from one regime to the other occurs at fixed values of $a$,$\theta$ when a critical value of the impact parameter $B$ is reached; the corresponding criterion is discussed in detail below.
In the mode of existence of three critical radii, the innermost solution is located in the close vicinity of the event horizon, the second inner solution is inside below the unstable spherical photon orbit region \cite{2003GReGr..35.1909T}, and the only external solution may be located far from the emitting surface (see Figs. 1 -- 3 in \cite{Bini2011} for details).
In the case studied here of a rigidly rotating emitting surface, the critical radii located inside the emitting surface are not relevant. In addition, the range of the impact parameter $b$ and thus the formal existence of the regime of three critical radii are limited by the maximum rotational frequency of a NS and, in the case of a BH, by the fact that the rotation of the emitting surface cannot be superluminal (see Secs. \ref{sec:brange_kerr} -- \ref{sec:NS_surface}).  To find the criterion for distinguishing the regimes with one critical radius and with three critical radii, we approximated Eq. (\ref{eq:crit_hyper2}) with a third order polynomial. Then, in order to count the number of solutions of Eq. (\ref{eq:crit_hyper2}), we exploited the standard formula for third order algebraic equation, which can be written in the canonical form as $a_3x^3+a_2x^2+a_1x+a_0=0$ \cite{1965hmfw.book.....A}. 
The discriminant $\Delta_{\rm III}$ is defined as follows
\begin{equation}
\label{eq:delta3}
\Delta_{\rm III}=4(3a_3a_1-a_2^2)^3+(27a_3^2a_0-9a_3a_2a_1+2a_2^3)^2.
\end{equation}
In the case of $\Delta_{\rm III}>0$ there exist three real solutions while in the case of $\Delta_{\rm III}<0$ there is only one real solution. In the limit case of $\Delta_{\rm III}=0$ the equation has a multiple real root. To this end, we consider separately the following two functions
\begin{eqnarray}  
y_1&&=a(n)^{\hat r}+2\theta(n)^{\hat r}{}_{\hat\varphi}\cos\beta+k_{\rm (Lie)}(n)^{\rm r}\cos^2\beta,\label{eq:func1}\\
y_2&&=\frac{A(1+ bN^\varphi)^2\sin^4\beta}{N^2 \sqrt{R_{b,q}(r}}.\label{eq:func2}
\end{eqnarray}

Let us consider first Eq. (\ref{eq:func1}), and its Taylor expansion for $1/r\to 0$; we obtain
\begin{equation}
\begin{aligned}
y_1&\approx\frac{r^2}{\rho\sqrt{\Sigma^5}}\left[r^3+r^2 \left(1-\frac{b^2}{\sin^2\theta}\right)\right.\\
&\left.+r \left(-6 a b+\frac{3 b^2}{\sin^2\theta}-a^2\cos2 \theta+2 a^2+3\right)\right.\\
&\left.+6 a b-\frac{\left(2 a^2 \cos 2 \theta +a^2+3\right) b^2}{2 \sin^2\theta}\right.\\
&\left.+\frac{1}{2} \left(3 a^2 \cos2 \theta-4 a^2+5\right)\right],
\end{aligned}
\end{equation}
where $\rho\sqrt{\Sigma^5}$ is a factor in common with all terms in $y_1$ (for further details see Table 1 in \cite{DeFalco20183D}) and can be approximated through the following polynomial
\begin{equation}
\begin{aligned}
\frac{1}{\rho\sqrt{\Sigma^5}}&\approx
-\frac{2a^2 \sin ^2\theta}{r^{10}}-\frac{a^2\left(5 \cos^2\theta+2\right)}{2 r^9}+\frac{1}{r^7}.
\end{aligned}
\end{equation} 
Combining these results, we get
\begin{eqnarray} \label{eq:approx1}
y_1&&\approx
\frac{1}{r^5}\left[r^3+r^2 \left(1-\frac{b^2}{\sin^2\theta}\right)\right.\\
&&\left.+r \left(-6 ab+\frac{3b^2}{\sin^2\theta}-\frac{5a^2\cos^2\theta}{2}-a^2\cos2 \theta+a^2+3\right)\right.\notag\\
&&\left.-\frac{a^2}{2}\left(5 \cos^2\theta+2\right) \left(1-\frac{b^2}{\sin^2\theta}\right)-2 a^2 \sin^2\theta+6ab\right.\notag\\
&&\left.\frac{1}{2} \left(3 a^2 \cos2 \theta-4 a^2+5\right)-\frac{\left(2 a^2 \cos2\theta+a^2+3\right)b^2}{2\sin^2\theta}\right]\notag
\end{eqnarray}

Let us now consider Eq. (\ref{eq:func2}), and its Taylor expansion for $1/r\to 0$
\begin{eqnarray} \label{eq:approx2}
y_2&&\approx
\frac{A}{r^5} \left\{r^3+2r^2+r\left[-\frac{3}{2} \frac{b^2}{\sin^2\theta}-\frac{a^2}{2}\left(\cos^2\theta+1\right)+4\right]\right.\notag\\
&&\left.\qquad\quad-2ab+2 a^2 \sin^2\theta-4 a^2+8\right\}.
\end{eqnarray}

Combining Eqs. (\ref{eq:approx1}) and (\ref{eq:approx2}) we obtain an approximation of Eq. (\ref{eq:crit_hyper2}) through a polynomial of third order of the form 
\begin{equation} \label{eq:3eq}
\frac{a_3r^3+a_2r^2+a_1r+a_0}{r^5}=0,
\end{equation}
where
\begin{eqnarray}
\label{eq:deltacoef}
a_3&&=1-A,\\
a_2&&=1-2 A-\frac{b^2}{\sin^2\theta},\notag\\
a_1&&=-\frac{A}{2} \left[-\frac{3b^2}{\sin^2\theta}-a^2(1+\cos ^2\theta)+8\right]-6 a b\notag\\
&& \frac{3b^2}{\sin^2\theta}-\frac{a^2}{2} \left(5 \cos ^2\theta+2\right)-a^2 \cos2\theta+2 a^2+3,\notag\\
a_0&&=-A \left(-2 a b+2 a^2 \sin ^2\theta-4 a^2+8\right)\notag\\
&&-\frac{a^2}{2} \left(5 \cos ^2\theta+2\right)\left(1-\frac{b^2}{\sin^2\theta}\right)+6 a b-2 a^2 \sin ^2\theta\notag\\
&&-\frac{\left(2 a^2 \cos2 \theta+a^2+3\right) b^2}{2\sin^2\theta}+\frac{\left(3 a^2 \cos2 \theta-4 a^2+5\right)}{2}.\notag
\end{eqnarray}
Therefore, the above coefficients determine the value of the discriminant (\ref{eq:delta3}).
Physically meaningful critical radii are located above the emitting surface.

\subsection{Constant latitude suspended orbits bound on the critical hypersurface}
\label{sec:susporbits}

In the simpler case of the test radiation field with zero angular momentum ($b=0$) investiged in the previous paper \cite{DeFalco20183D}, the particle captured on an off-equatorial circular orbit bound on the critical hypersurface with a constant polar coordinate $\bar{\theta}$ must fulfil the condition $\cos\psi=0, \frac{d\psi}{d\tau}=0,\nu=0 $ for which a complete balance of all forces acting at the critical hypersurface is attained. Here we investigate the condition for similar suspended off-equatorial circular orbits for the more complex case of the rotating test radiation field ($b\ne 0$).

For a particle to be captured on an suspended  off-equatorial circular orbit bound on the critical hypersurface, i.e. without the presence of the effect of latitudinal drift, the condition $d\psi/d\tau=0$ must be fulfilled along with the solution of Eq. (\ref{eq:crit_hyper2}).
The test particle velocity on the critical hypersurface is equal to the azimuthal photon velocity ($\nu=\cos\beta$). Using Eq. (\ref{EoM2}) with $\alpha=0$, $r=r_{\rm crit}(\bar{\theta})$, $\theta=\bar{\theta}$, imposing $d\psi/d\tau=0$, we obtain
\begin{equation}
\begin{aligned}\label{eq:susporbit}
&\gamma\nu\left[a(n)^{\hat \theta}+k_{\rm (Lie)}(n)^{\hat \theta}\,\nu^2+2\nu\sin^2\psi\ \theta(n)^{\hat \theta}{}_{\hat \varphi}\right]+\\
&+\frac{A(1+bN^\varphi)^2\nu\cos\psi(1-\nu^2\sin\psi)}{N^2\sqrt{R_{b,q}(r_{\rm crit})}}=0.
\end{aligned}
\end{equation}
By solving this implicit equation for $\psi$, we obtain the condition for suspended circular orbits.  \footnote{Note that in the case reported in the previous paper \cite{DeFalco20183D}, i.e. for $b=0$, the condition for suspended circular orbits takes the very simple form $a(n)^{\hat \theta}=0$.}.
\begin{figure}[h!]
	\centering
		\includegraphics[width=1.0\linewidth]{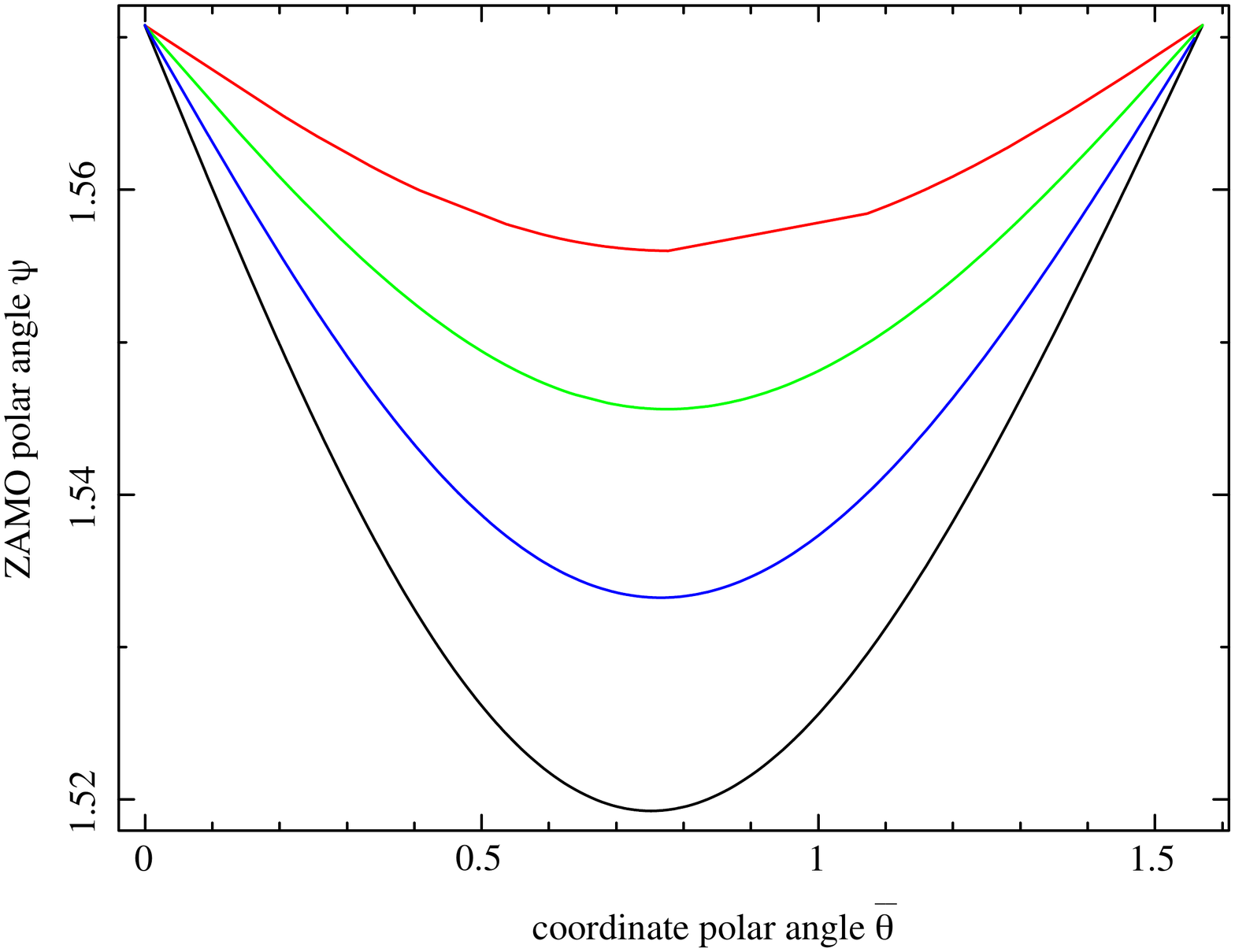}
			\includegraphics[width=1.0\linewidth]{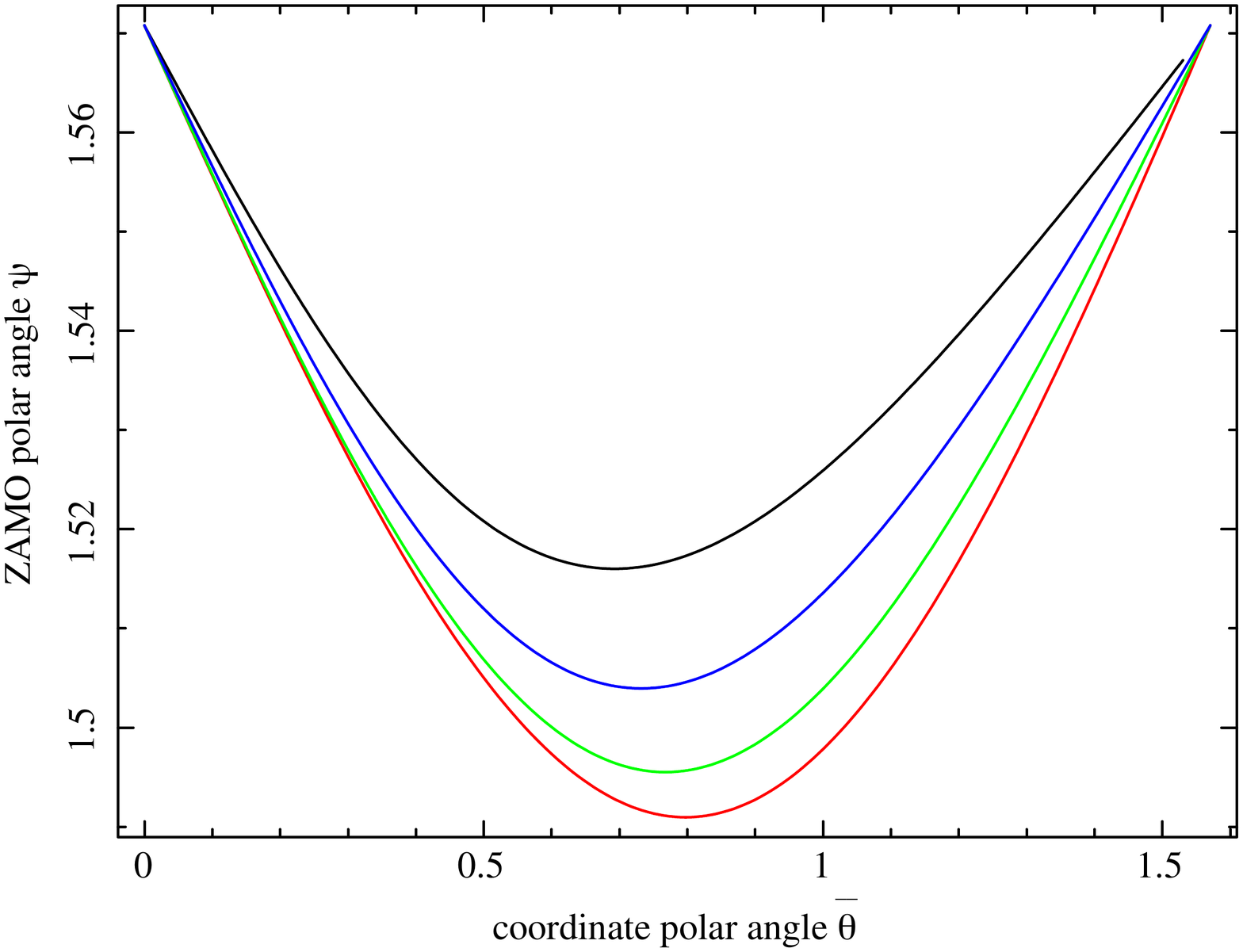}
		\caption{The local polar angle in the ZAMO frame $\psi$ as a function of the latitudinal coordinate $\bar{\theta}$ of circular suspended orbits in the case of NS of mass $M=1.5M_\odot$ and radius $R_\star=6M$. Top panel: The plot is constructed for fixed luminosity parameter $A=0.8$ and star rotation frequency $f_{\star}=300$ Hz (red), $f_{\star}=400$ Hz (green), $f_{\star}=500 $Hz (blue), $f_{\star}=600$ Hz (black). Bottom panel: The plot is constructed for fixed star rotation frequency $f_{\star}=700$ Hz and luminosity parameter $A=0.7$ (black), $A=0.7$ (blue), $A=0.8$ (green), $A=0.85$ (red).}
	\label{fig:FigSO}
\end{figure}

The condition (\ref{eq:susporbit}) significantly depends on the radius of the emitting surface $R_\star$ and its rotational frequency $\Omega_\star$. In Fig \ref{fig:FigSO}, we illustrate the behaviour of the local polar angle in the ZAMO frame $\psi$ 
as a function of the latitudinal coordinate $\bar{\theta}$ of circular suspended orbits bound on the critical hypersurface. 
On the equator, the value of the local polar angle is $\psi=\pi/2$ due to the mirror symmetry.  The angle $\psi$ decreases with growing latitudinal coordinate $\bar{\theta}$ to its minimum, then $\psi$ grows and it reaches the value of $\psi=\pi/2$ again on the polar axis, where the test radiating field is non-rotating ($b=0$).

\subsection{Critical hypersurfaces in the case of a NS}
\label{sec:CHNS}
\begin{figure}[th!]
	\centering
		\includegraphics[trim=0.5cm 0cm 0cm 0cm,scale=0.32]{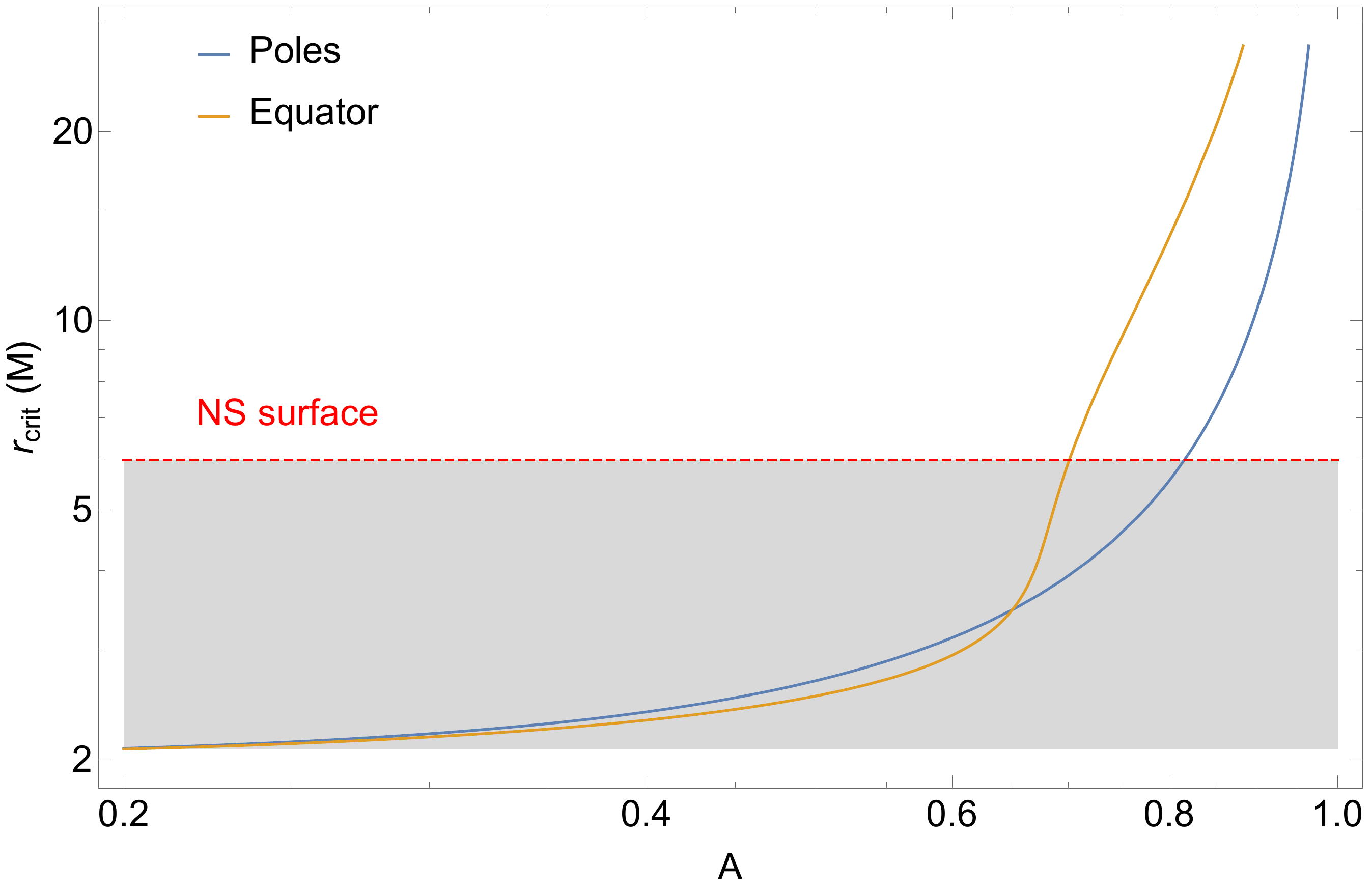}
	\caption{Critical radius $r_{\rm (crit)}$ at the poles (blue line) and at equator (orange line) as a function of the luminosity parameter $A$ for a NS model with rotation frequency $f_{\star}=600$, $\Omega_\star=0.031M^{-1}\,,a=0.41$). The dashed red line represents the NS surface, the gray area indicates the (nonphysical) solutions inside the NS.}
	\label{fig:Fig2}
\end{figure}
\begin{figure}[th!]
	\centering
		\includegraphics[trim=0.5cm 0cm 0cm 0cm,scale=0.35]{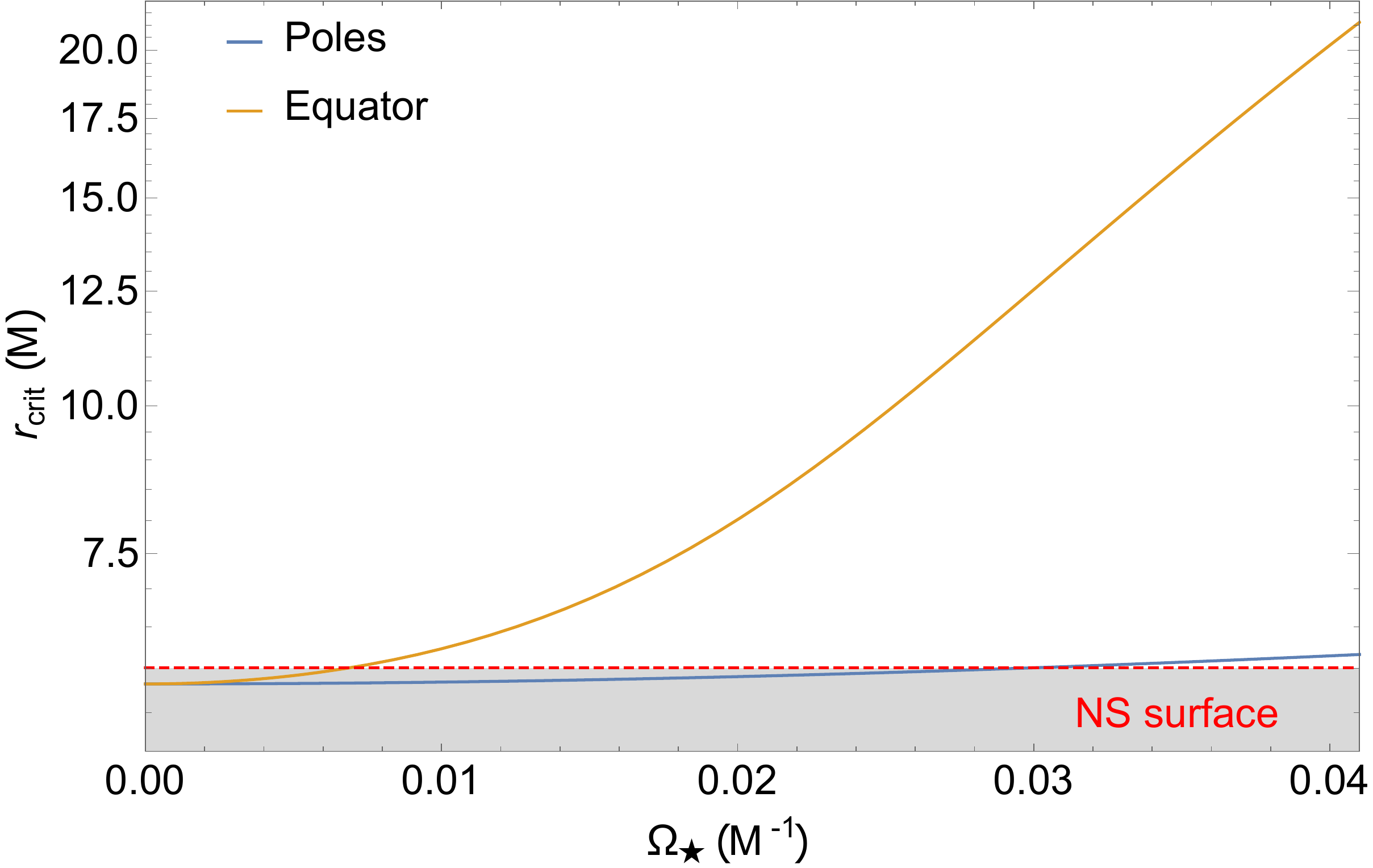}
	\caption{Critical radius $r_{\rm (crit)}$ at the poles (blue line) and at the equator of the NS (orange line) as a function of the NS rotation angular velocity $\Omega_\star$ for a luminosity parameter of $A=0.81$. The dashed red line represents the NS surface, the gray area indicates the (nonphysical) solutions inside NS.}
	\label{fig:Fig3}
\end{figure}

\begin{figure*}[th!]
	\centering
	\hbox{
		\includegraphics[scale=0.28]{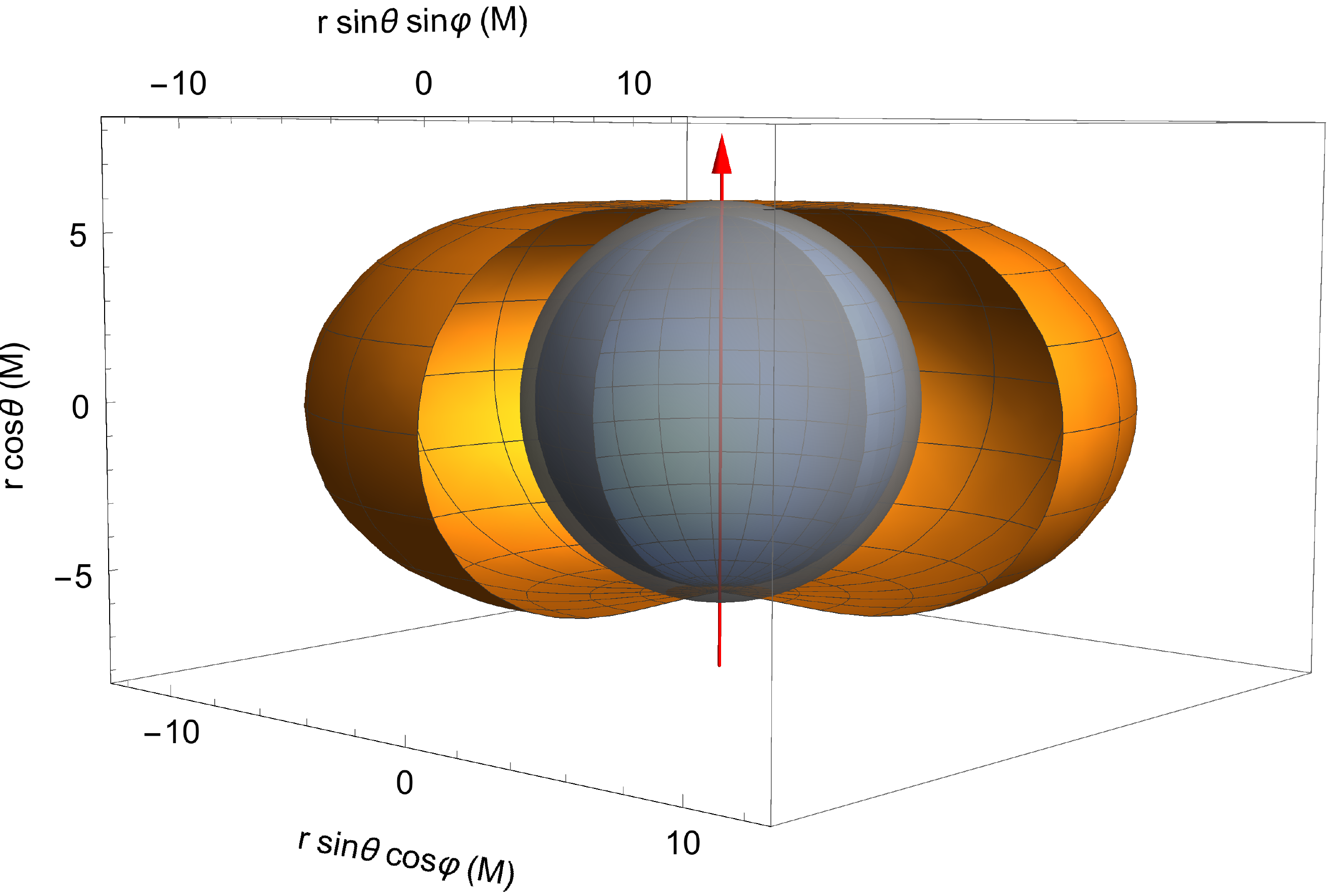}
		\hspace{0.3cm}
		\includegraphics[scale=0.36]{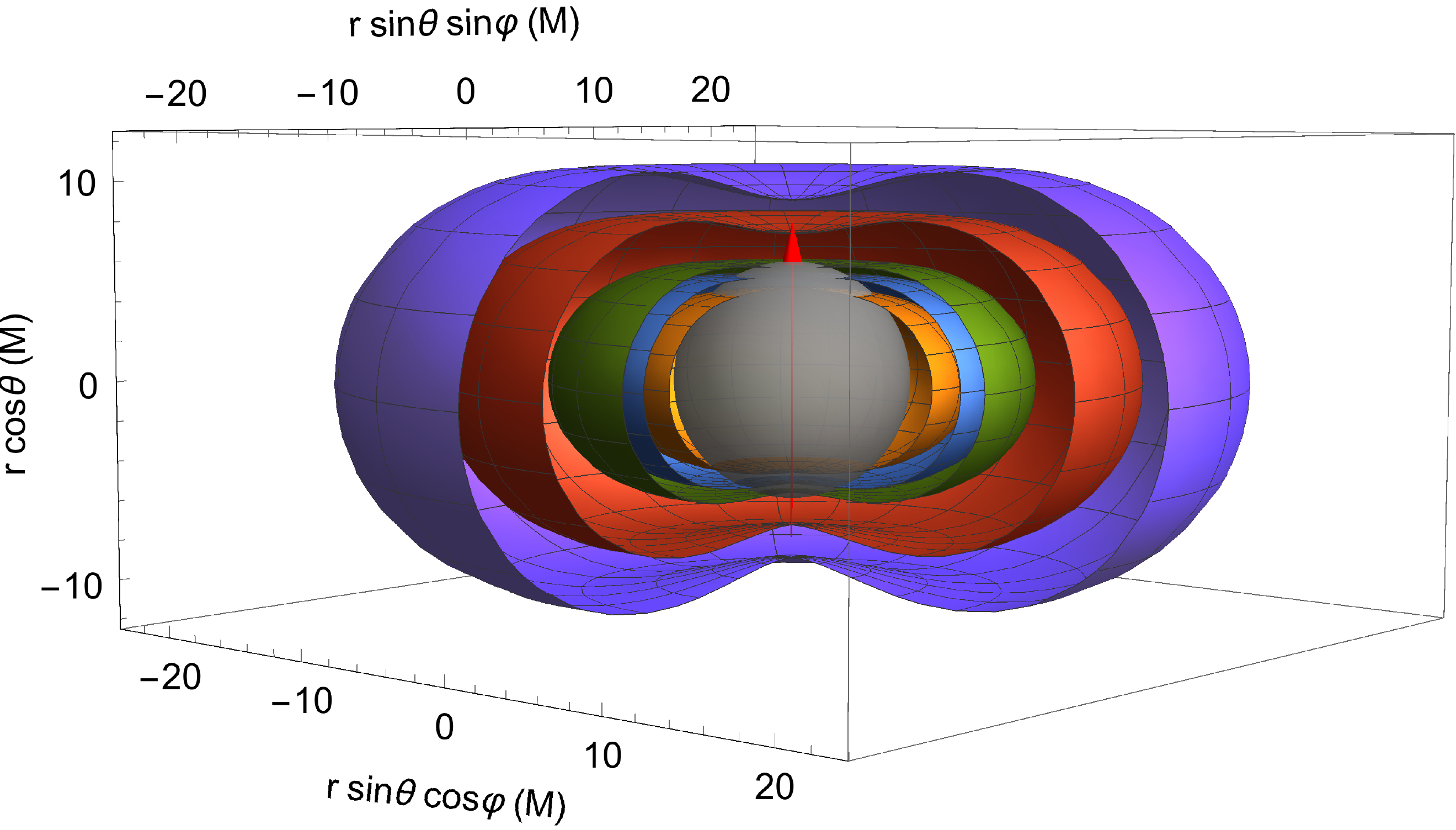}}
	\caption{Left panel: Critical hypersurfaces for the case of a rotating NS with Kerr parameter $a=0.41$ (orange, $f_{\star}=600$ Hz, $\Omega_\star=0.031M^{-1}$) and for the non-rotating case (blue). For a non-rotating  NS, the critical radius is constant and it takes unphysical value of $r_{\rm (crit)} \sim 5.56M$. In the case of a rotating NS, the critical radius takes the value $r^{\rm eq}_{\rm(crit)} \sim 12.05M$ in the equatorial plane, while it takes the value of  $r^{\rm pole}_{\rm(crit)} \sim 5.74M$ at the poles. The relative luminosity of the radiating field is fixed at $A=0.8$. Right panel: Critical hypersurfaces for a relative luminosity $A=0.75,\, 0.78, \,0.8, \,0.85, \,0.88$ at Kerr parameter $a=0.41$ ($f_{\star}=600$ Hz, $\Omega_\star=0.031M^{-1}$). The corresponding critical radii in the equatorial plane are $r^{\rm eq}_{\rm(crit)} \sim 8.88M,\  10.61M,\ 12.05M,\ 17.26M,\ 22.43M\ $, while at poles they are $r^{\rm pole}_{\rm(crit)} \sim 4.73M,\  5.28M,\  5.74M,\  7.43M,\  9.11M$. In both panels the gray sphere represents the NS surface, separating the physical solutions (outside its surface) from the unphysical ones (inside its surface). The red arrow is the polar axis.}
	\label{fig:Fig4}
\end{figure*}

\begin{figure*}[th!]
	\centering
	\hbox{
		\includegraphics[trim=5cm 6.2cm 1cm 0cm,scale=0.35]{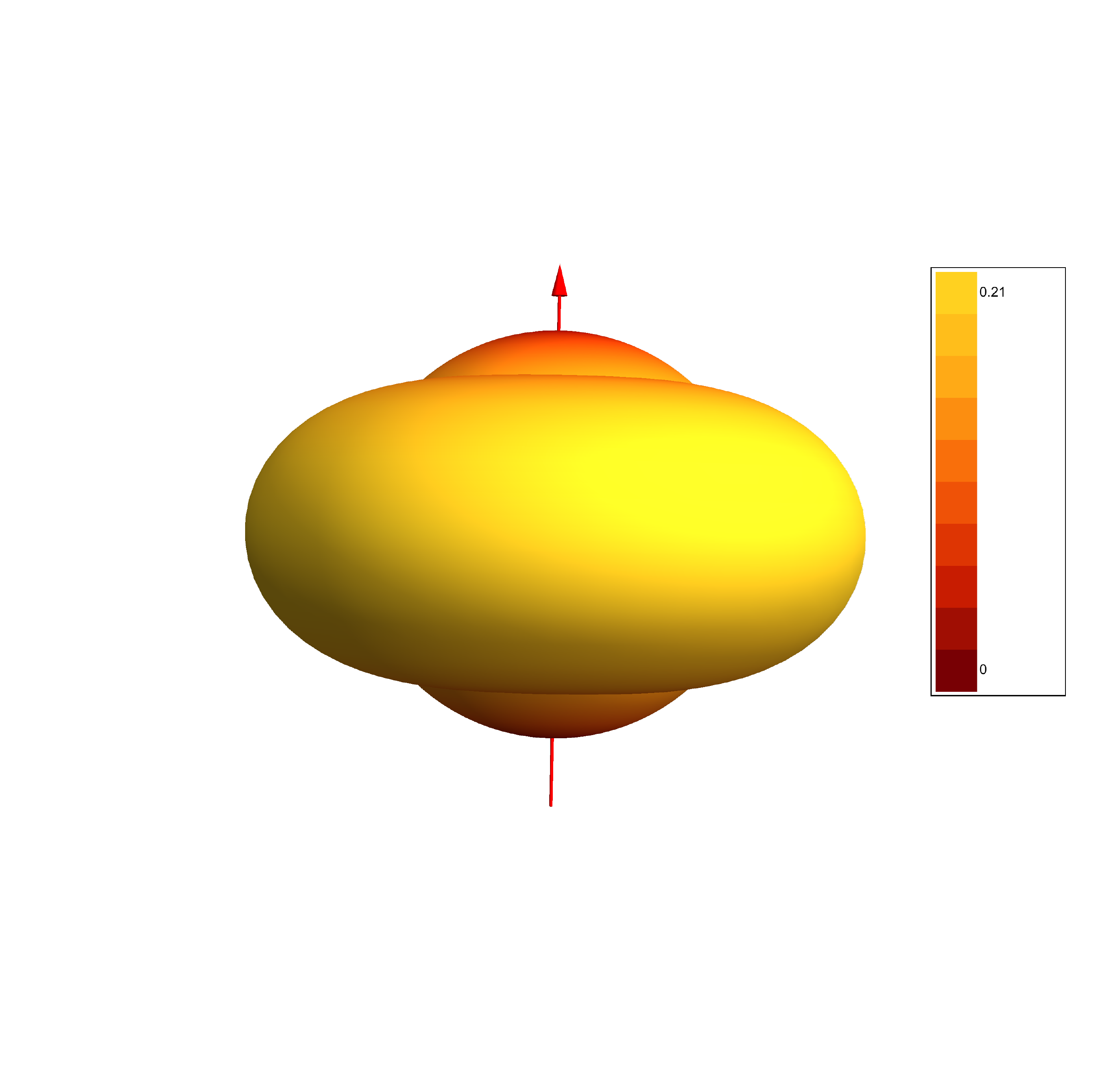}
		\hspace{0.3cm}
		\includegraphics[scale=0.33]{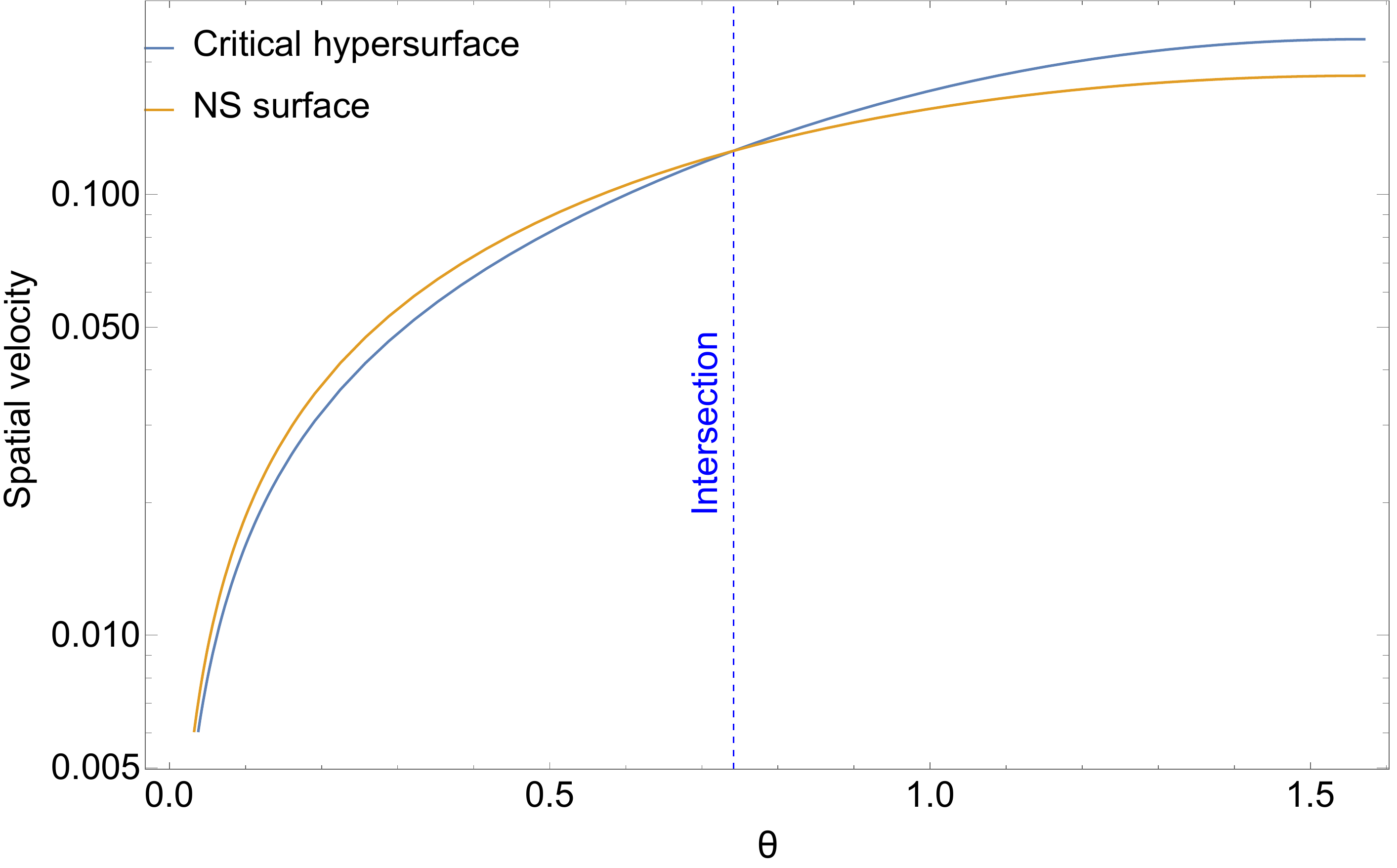}}
	\caption{Left panel: The spatial velocity of test particles $v_{\rm (crit)}$ captured on the critical hypersurface forming a lobe and the spatial velocity of the NS surface $v_{\rm NS}$ as measured by a static observer at infinity as functions of polar angle $\theta$.  The spatial velocity values ranges between zero and maximum value of $0.21$. The red arrow represents the polar axis. For the critical hypersurface size dimensions see the corresponding case in Fig. \ref{fig:Fig4}. Right panel: The spatial velocity of the critical hypersurface forming a lobe $v_{\rm (crit)}$ (blue) and the spatial velocity of NS surface $v_{\rm NS}$ (yellow) as a function of the polar angle $\theta$. The vertical dashed blue line is located at $\theta=42.5^\circ$ representing the polar angle at which the critical hypersurface lobe intersects the NS surface. Both plots are constructed for template NS with Kerr parameter $a=0.41$ ($f_{\star}=600$ Hz, $\Omega_\star=0.031M^{-1}$) and for a radiating field with luminosity parameter $A=0.75$.}
	\label{fig:Fig4bis}
\end{figure*}
\begin{figure}[th!]
	\centering
			\includegraphics[width=1.0\linewidth]{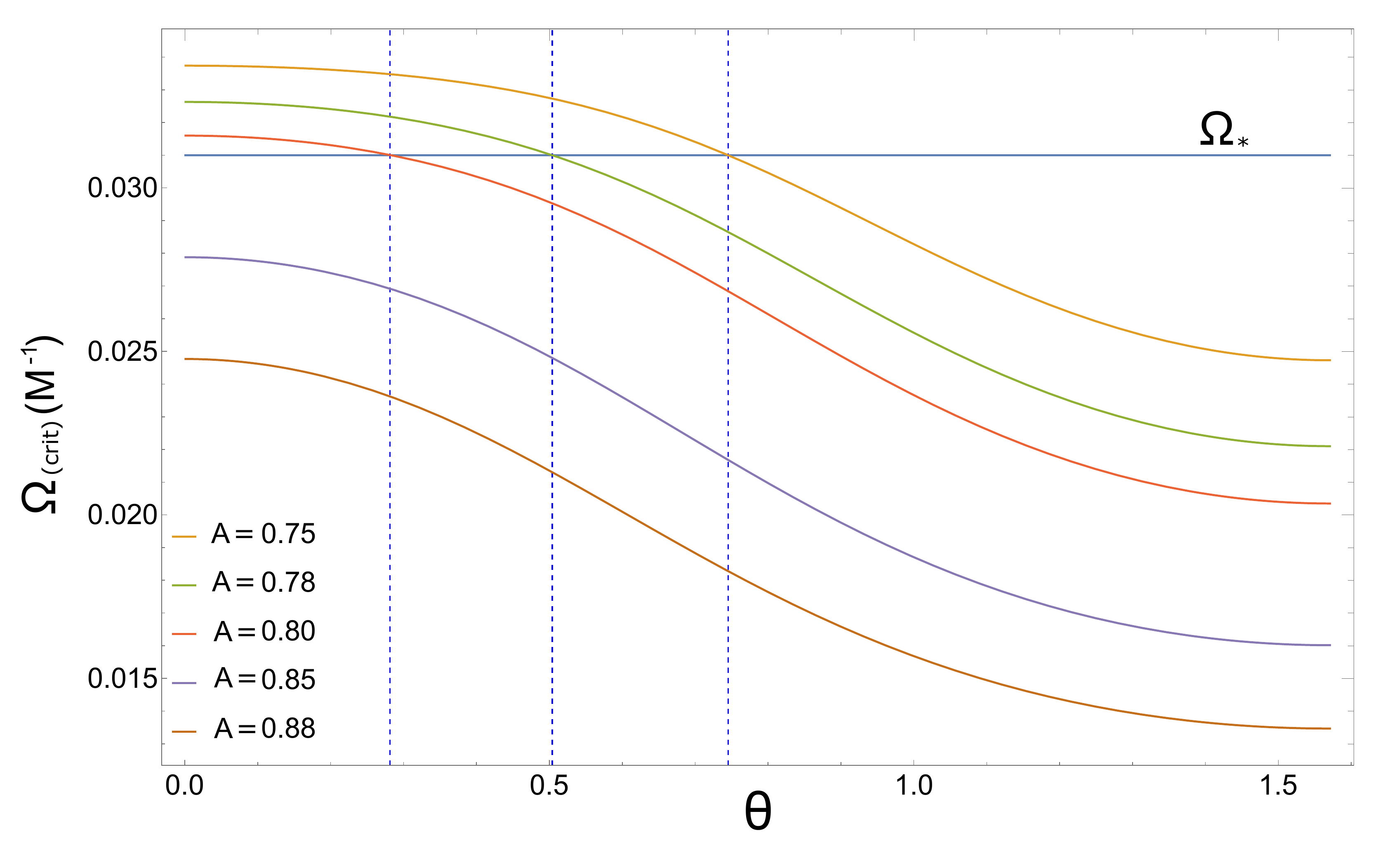}
	\caption{The angular velocity $\Omega_{\rm (crit)}$ of test particles captured on the critical hypersurface forming a lobe as a function of polar angle $\theta$.  The vertical dashed blue line is located at $\theta=42.5^\circ$ representing the polar angle at which the critical hypersurface lobe intersects the NS surface. The plot is constructed for a NS with Kerr parameter $a=0.41$ ($f_{\star}=600$ Hz, $\Omega_\star=0.031M^{-1}$) and for radiating field with luminosity parameter $A=0.75$.}
	\label{fig:Fig4bis1}
\end{figure}

We consider a template NS of mass $M=1.5M_\odot$ and radius $R_\star=6M$. The emitting source we consider is the spherical surface of the NS. All solutions lying inside the NS have no physical meaning (these include also the internal solutions Eq. (\ref{eq:crit_hyper2}) in the close vicinity of the event horizon and in the region of the spherical photon orbits).  

In Fig. \ref{fig:Fig2} we investigate the way the critical radius changes along the polar axis and at the equator of NS  for different values of the luminosity parameter. The NS rotation frequency is fixed here at  $f_\star=600$ Hz (corresponding to $\Omega_\star=0.031M^{-1}$ and through Eq. (\ref{eq:NSa}) to Kerr parameter $a=0.41$). The critical hypersurface above the emitting surface is always oblate (note that it changes from prolate to oblate for $A\simeq 0.65$ in the non-physical region inside the NS) . Moreover, the difference of the polar and equatorial value of the critical radius gives rise to an interesting new effect. For the luminosity range $A \in[0.705,0.81]$, the physically meaningful part of the critical hypersurface located above the surface of NS forms a lobe around the star equator (see the right panel of Fig \ref{fig:Fig4} and Fig. \ref{fig:Fig4bis}). For $A<0.705$, the entire critical hypersurface is located inside NS (and thus is nonphysical), while for $A>0.81$, the entire critical hypersurface is located above the  NS surface. For the lower limiting value $A=0.705$, the critical hypersurface just emerges from the NS interior at the equator, while for the upper limiting value $A=0.81$, the completely emerged critical hypersurface touches only the poles of NS.

Figure \ref{fig:Fig3} shows the way the critical radius behaves in terms of the NS angular velocity. The luminosity parameter is fixed at the value of $A=0.81$. The  NS rotation frequency range is 
$f_\star \in[0,800]$ Hz (corresponding to $\Omega_\star\in[0,0.041]M^{-1}$ or $a\in[0,0.55]$, note that fastest know NS spins at $\sim716$ Hz \cite{Hessels2006}). We consider only positive rotation frequencies, because $\Omega_\star$ is linearly connected to the Kerr parameter  $a$ by the relation (\ref{eq:NSa}).
On the poles, where surface velocity is zero and frame dragging effect is absent, the critical radius value varies slightly only as a result of changes in spacetime geometry as determined by $a$ growing with the $\Omega_\star$. The critical radius on the pole reaches the radius of the NS for $\Omega_\star=0.031M^{-1}$, which is in accordance with the profile of polar critical radius plotted in Fig. \ref{fig:Fig2}. On the contrary there is a much stronger dependence on $\Omega_\star$ at the equator because the surface velocity and magnitude of frame dragging reach their maxima. Thus, similar to Fig. \ref{fig:Fig2}, the critical hypersurface located above the emitting surface of NS is always oblate.
 
The left panel of Fig. \ref{fig:Fig4} compares the shape of the critical hypersurfaces for a rotating template NS with Kerr parameter $a=0.41$ ($f_\star=600$ Hz, $\Omega_\star=0.031M^{-1}$)
and for a non-rotating NS ($a=0$), where the luminosity parameter of the radiating field is fixed at the value of $A=0.8$. In the rotating case, the critical radius is $r_{\rm crit}^{\rm eq}\sim 12.05M$ at the equator and $r_{\rm crit}^{\rm pole}\sim 5.74M$ at the poles. In the non-rotating case, the critical hypersurface naturally takes the shape of a sphere with the radius $r_{\rm crit}\sim 5.56M$. Indeed, the non-rotating case has no physical meaning as the resulting equilibrium sphere is located inside NS. However, such a non-rotating solution coincides with the case of a pure radially moving test field photon on the background of the Schwarzschild spacetime geometry discussed in \cite{DeFalco20183D}, where we calculated the critical radius as a result of a static equilibrium.

The right panel of Fig. \ref{fig:Fig4} illustrates the shape of the critical hypersurfaces for the values of the relative luminosity in the interval $0.75-0.88$ and for a constant value of the Kerr parameter $a=0.41$ ($f_{\star}=600$ Hz, $\Omega_\star=0.031M^{-1}$). The figure clearly demonstrates that in the appropriate luminosity range, the critical hypersurface forms only a lobe around the NS equator .
In Fig. \ref{fig:Fig4bis} we plot the spatial velocity of  the test particle captured on the critical hypersurface $v_{\rm (crit)}$ as measured by a static observer at infinity, for a luminosity parameter of $A=0.75$ and a NS Kerr parameter of $a=0.41$ ($f_{\star}=600$ Hz, $\Omega_\star=0.031M^{-1}$). The lobe of the critical hypersurface intersects the NS surface at $\theta=42.5^\circ$. 

The spatial velocity of test particles captured on the critical hypersurface is
\begin{equation}
v_{\rm (crit)}=r_{\rm (crit)}\sin\theta\,\Omega_{\rm (crit)}. 
\end{equation}
By using Eq. (\ref{four_velocity}) and Eq. (\ref{eq:crit_veloc}), the angular velocity of test particles captured on the critical hypersurface is 
\begin{equation}
\Omega_{\rm (crit)}=U^\phi/U^t\,=\frac{N\cos\beta}{\sqrt{g_{\varphi\varphi}}}+\Omega_{\rm ZAMO}\,.
\end{equation}
The spatial velocity of the NS surface $\nu_{\rm NS}$ as measured by a static observer at infinity is\begin{equation}
v_{\rm NS}=R_\star\sin\theta\ \Omega_\star.
\end{equation}

The angular velocity of the particles captured on the critical hypersurface $\Omega_{\rm (crit)}$, which is equal to the angular velocity of NS surface $\Omega_\star$ at the intersection ring, decreases for increasing polar angle $\theta$, reaching the minimum value on the equatorial plane.
At the intersection ring, the spatial velocity of the NS surface $v_{\rm NS}$ and that of test particles captured on the critical hypersurface $v_{\rm (crit)}$ are also equal. However $v_{\rm (crit)}$ is always higher than the $v_{\rm NS}$ on the critical hypersurface lobe, owing to the growth of $r_{\rm (crit)}$ with the polar angle $\theta$. $v_{\rm (crit)}$ also grows with $\theta$, and it reaches its maximum value $v_{\rm (crit)}(\theta=\pi/2)=0.21$ on the equatorial plane.

\subsubsection{Examples of test particle orbits in the vicinity of the critical hypersurface around NS}
\begin{figure*}[th!]
\centering
\hspace{1cm}
\vbox{\hbox{\includegraphics[scale=0.29]{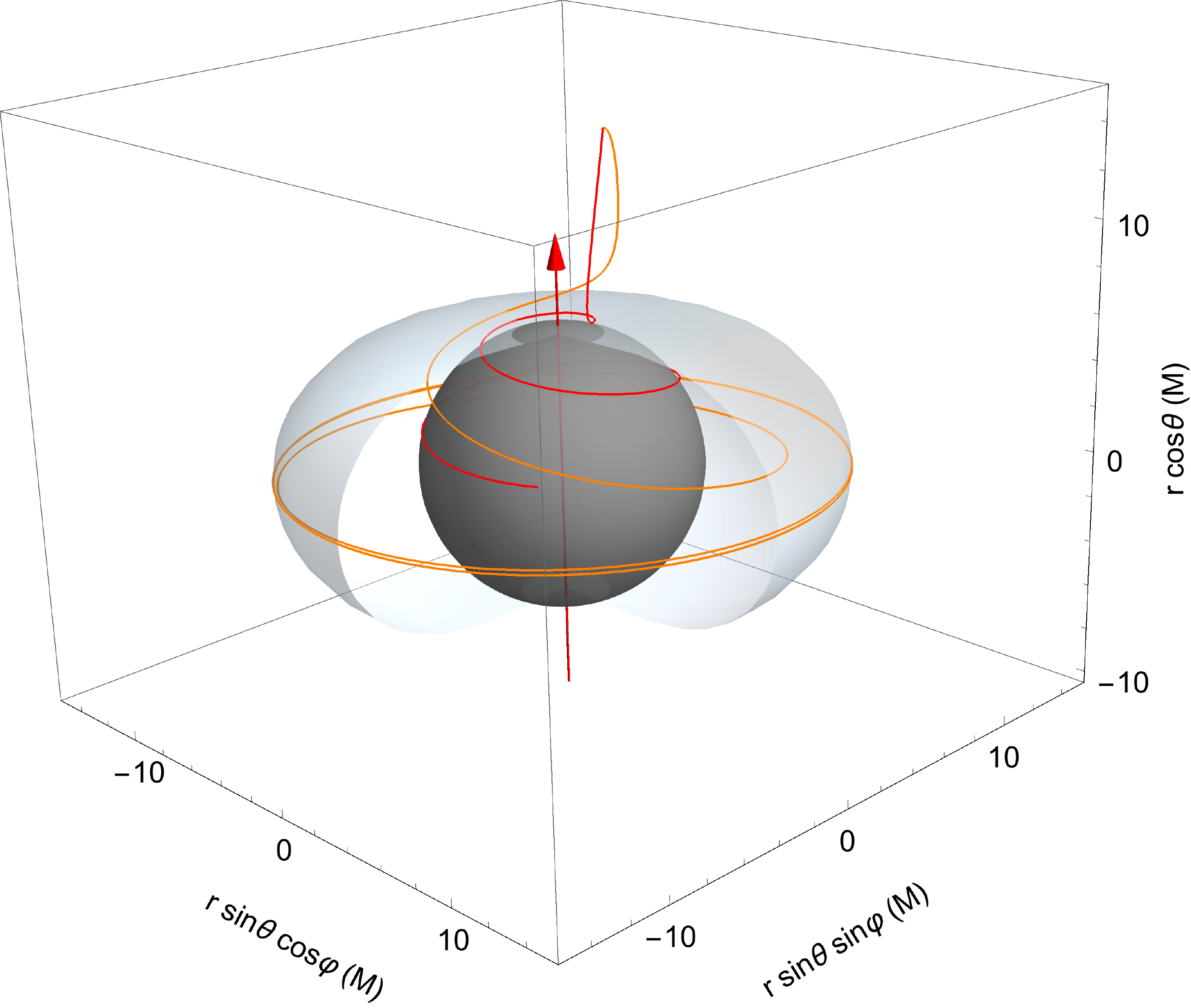}
	 \hspace{1cm}
	 \includegraphics[scale=0.22]{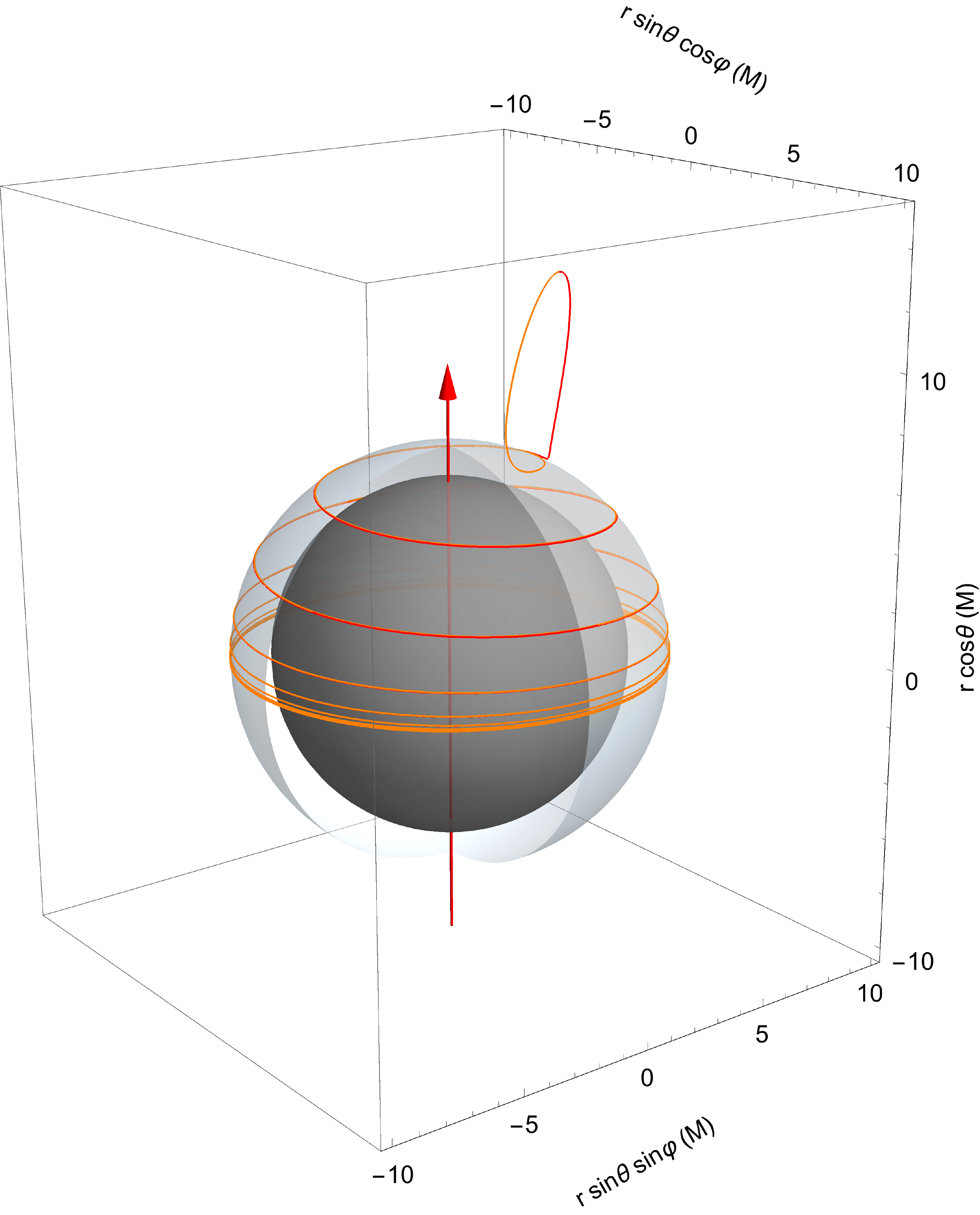}}
	 \vspace{0.5cm}
	\hbox{\includegraphics[scale=0.29]{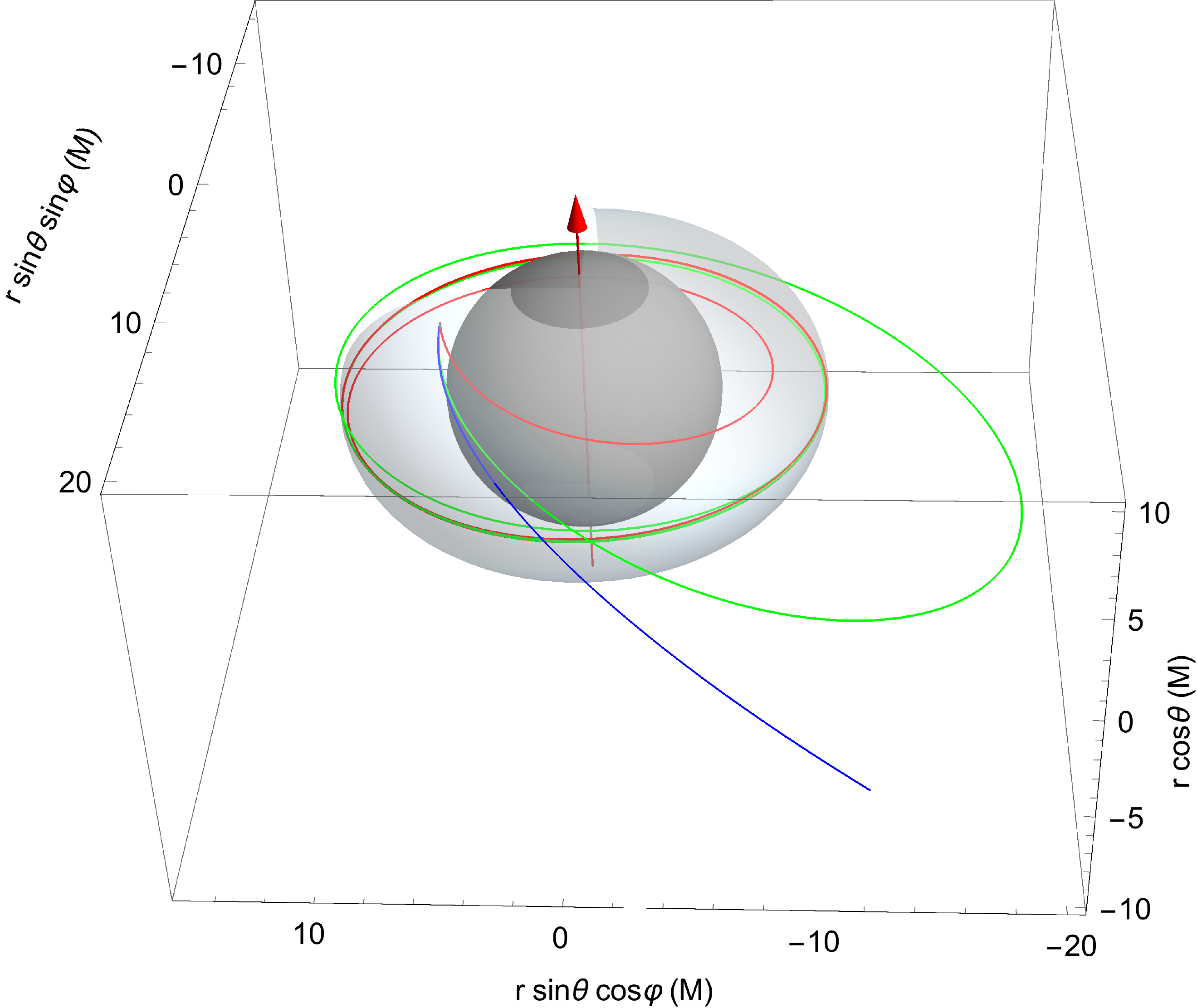}
	\hspace{1cm}
	\includegraphics[scale=0.21]{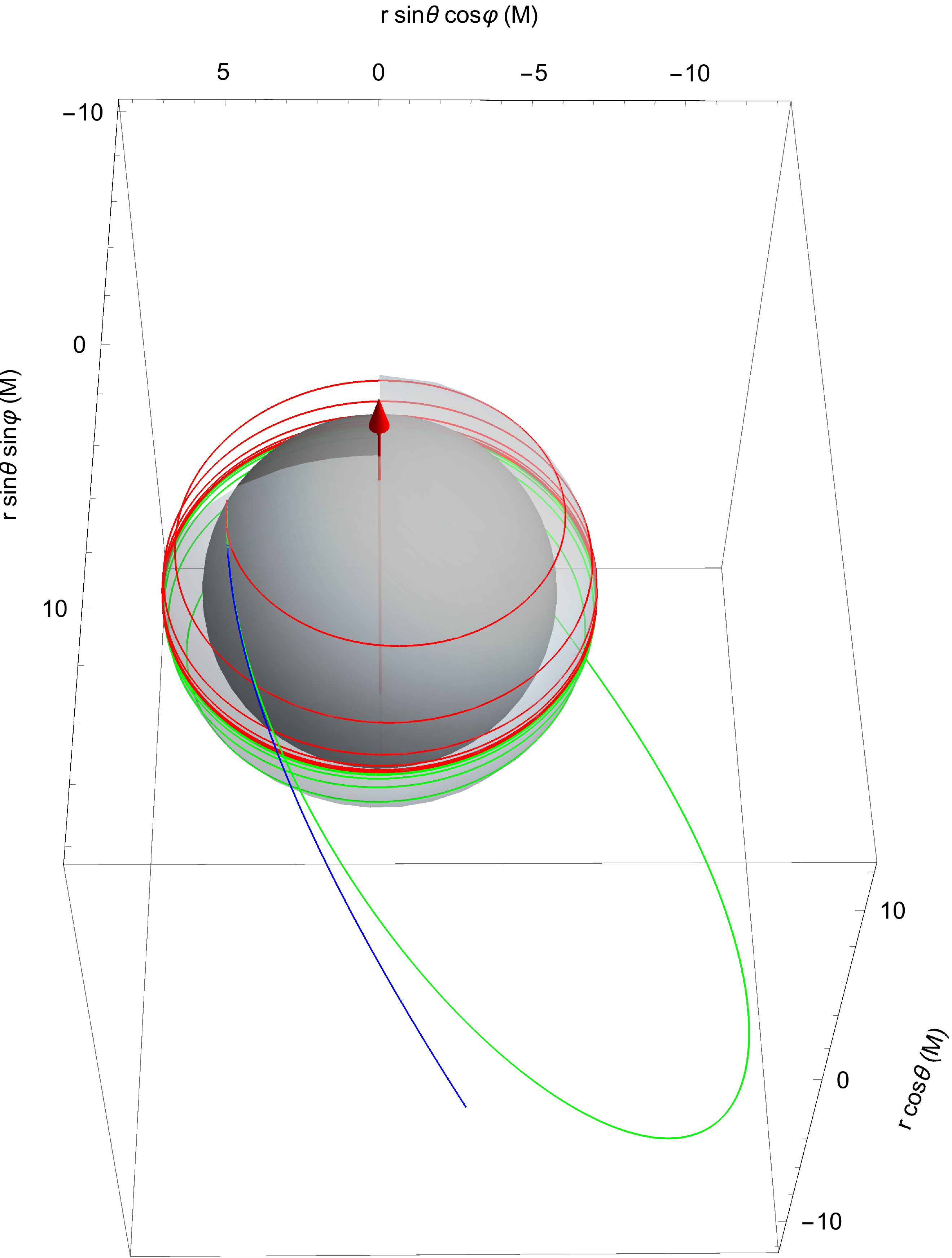}}}
\caption{Examples of trajectories of test particles interacting with the radiation field emitted by the surface of a slowly or rapidly rotating NS. Left panels: The cases of rapidly rotating template NS with Kerr parameter $a=0.41$ ($f_{\star}=600$ Hz, $\Omega_\star=0.031M^{-1}$) and with relative luminosity $A=0.8$. The critical hypersurface forms a big lobe. Right panels: The cases of slowly rotating template NS with Kerr parameter $a=0.07$ ($f_{\star}=100$ Hz, $\Omega_\star=0.005M^{-1}$) and with relative luminosity $A=0.85$. The critical hypersurface wraps the entire NS surface. Top panels: The test particles are emitted outside the critical hypersurface at $r_0=15M,\,\theta_0=10^\circ$ with the small initial velocity $\nu_0=0.01$ oriented in the azimuthal corotating direction (orange) and oriented radially towards the emitting surface (red). 
Bottom panels: The test particles are emitted inside the critical hypersurface at $r_0=7M$ with the initial velocity oriented in the azimuthal corotating direction.
Bottom left panel: The initial polar angle is $\theta_0=45^\circ$,  the magnitude of initial velocity is $\nu_0=0.01$ (red), $\nu_0=0.3$ (green), $\nu_0=0.38$ (blue). Bottom right panel: The initial polar angle is $\theta_0=60^\circ$,  the magnitude of initial velocity is $\nu_0=0.01$ (red), $\nu_0=0.36$ (green), $\nu_0=0.42$ (blue). The black sphere corresponds to the emitting surface of the NS. The blue-gray quasi-ellipsoidal oblate surface denotes the position of the critical hypersurface. }
	\label{fig:FigNSorbits}
\end{figure*}

In Fig. \ref{fig:FigNSorbits}, we compare the results of the integration of the trajectories of test particles interacting with a radiation field emitted by the surface of a slowly or rapidly rotating NS. The left panels of Fig \ref{fig:FigNSorbits} corresponds to the case of a rapidly rotating template NS with Kerr parameter $a=0.41$ ($f_{\star}=600$ Hz, $\Omega_\star=0.031M^{-1}$) and with relative luminosity $A=0.8$. The critical hypersurface forms a large flattened lobe, with $r_{\rm crit}^{\rm eq}\sim 12.05M$ at the equator and $r_{\rm crit}^{\rm pole}\sim 5.74M$ at the poles. The right panels of Fig \ref{fig:FigNSorbits} corresponds to the case of a slowly rotating template NS with Kerr parameter $a=0.07$ ($f_{\star}=100$ Hz, $\Omega_\star=0.005M^{-1}$) and relative luminosity $A=0.85$. The critical hypersurface envelopes the entire NS surface, its radius being $r_{\rm crit}^{\rm eq}\sim 7.37M$ at the equator and $r_{\rm crit}^{\rm pole}\sim 7.21M$ at the poles.

On the both top panels of Fig \ref{fig:FigNSorbits}, the test particles are emitted outside the critical hypersurface at $r_0=15M,\,\theta_0=10^\circ$ with a small initial velocity $\nu_0=0.01$ oriented in the azimuthal corotating direction (orange) and oriented radially towards the emitting surface (red). On the top left panel corresponding to the case of the rapid rotation, 
the trajectory of the particle with the azimuthally-oriented initial velocity (orange) intersects the critical hypersurface from the outside. The particle is then captured on the critical hypersurface after passing a radial turning point. After a very short latitudinal drift, the particle's trajectory is stabilized in the equatorial plane. On the contrary, the particle with radially oriented initial velocity falls down the NS surface. The top right panel of Fig \ref{fig:FigNSorbits} shows to the case of a slowly rotating template NS with Kerr parameter $a=0.07$ ($f_{\star}=100$ Hz, $\Omega_\star=0.005M^{-1}$) and relative luminosity $A=0.85$. The critical hypersurface envelopes the entire NS surface, its radius being $r_{\rm crit}^{\rm eq}\sim 7.37M$ at the equator and $r_{\rm crit}^{\rm pole}\sim 7.21M$ at the poles. In this case, the trajectories differ only in their initial behaviour. Both test particles are then captured on the critical hypersurface and drift latitudinally toward the equatorial plane, where a final, circular orbit is attained. 

On the both bottom panels of Fig \ref{fig:FigNSorbits}, the test particles  are emitted inside the critical hypersurface at $r_0=7M$ with the initial velocity $\nu$ oriented in the azimuthal corotating direction.
The behaviour of the test particles is qualitatively the same regardless of the velocity of NS rotation. The test particles emitted with sufficiently small velocity $\nu$ are captured on the critical hypersurface from the inside (red). The test particles emitted with greater velocity $\nu$ intersect the critical hypersurface from the inside and they also intersect the equatorial plane. The particle is then captured on the critical hypersurface after passing a radial turning point (green). Captured test particles drift latitudinally toward the equatorial plane, where a final, circular orbit is attained. Similarly as in the cases illustrated in the top panels, the latitudinal drift is slower in the case of the slow rotation. Finally, the test particles emmited with sufficiently high velocity $\nu$ intersect the critical hypersurface from the inside and then escape to infinity (blue).    

\subsection{Critical hypersurfaces in the case of a BH}
\label{sec:CHBH}

We consider a BH of mass $M=5M_\odot$ and spin $a=0.9$. We placed the model emitting surface at $R_\star=2.5M$.  The emitting surface rotates with angular velocity $\Omega_\star$, which is unrelated to the BH spin. In the case of this relatively high value of the spin and at the value of the angular velocity of the emitting surface $\Omega_{\star} \leq \Omega_{+}$, Eq. (\ref{eq:crit_hyper2}) has only one solution, and the BH is enveloped by one critical hypersurface at the most (see the condition (\ref{eq:delta3})). 

\begin{figure*}[th!]
	\centering
	\hbox{
		\includegraphics[scale=0.315]{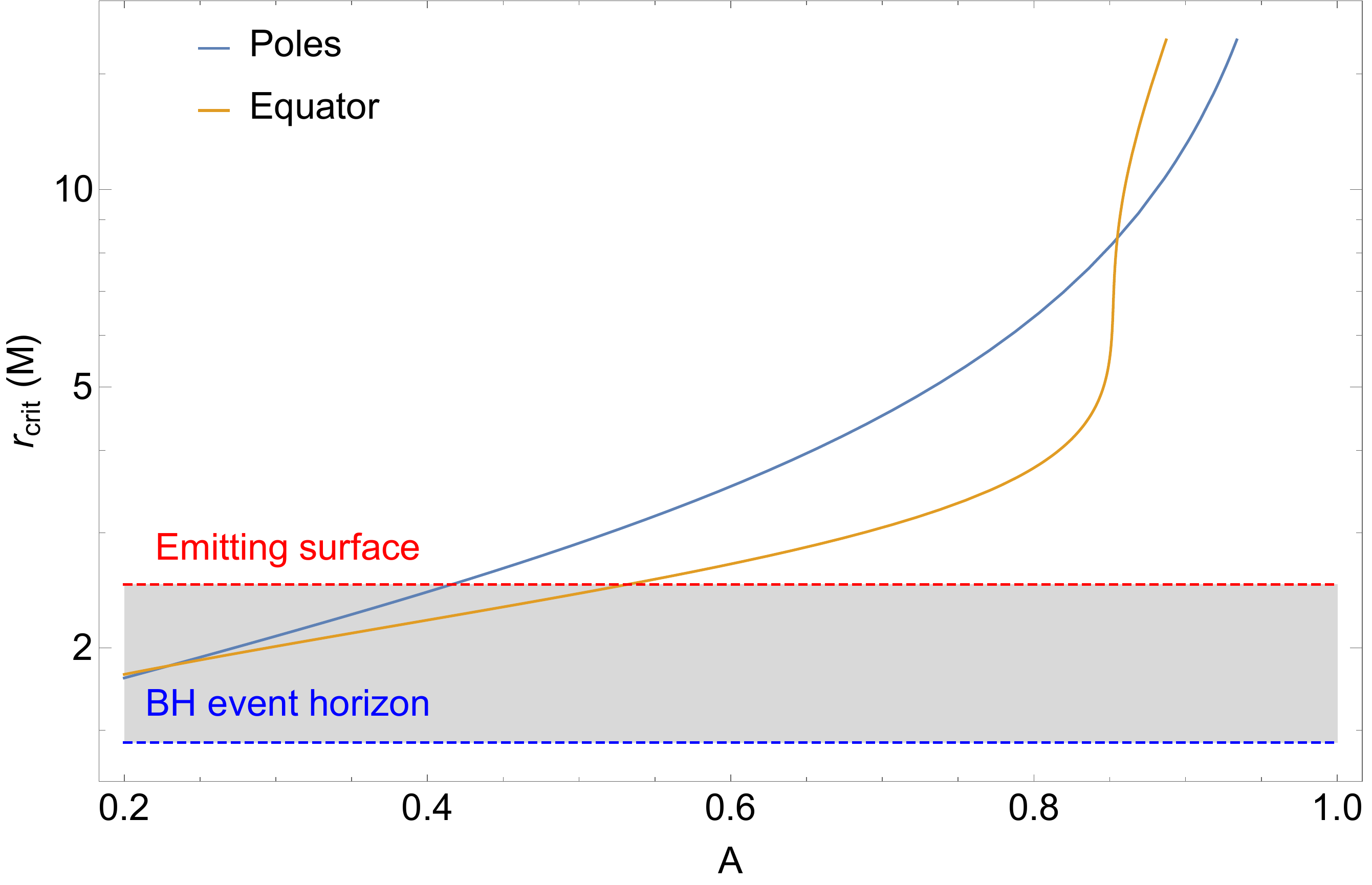}
		\hspace{0.3cm}
		\includegraphics[scale=0.32]{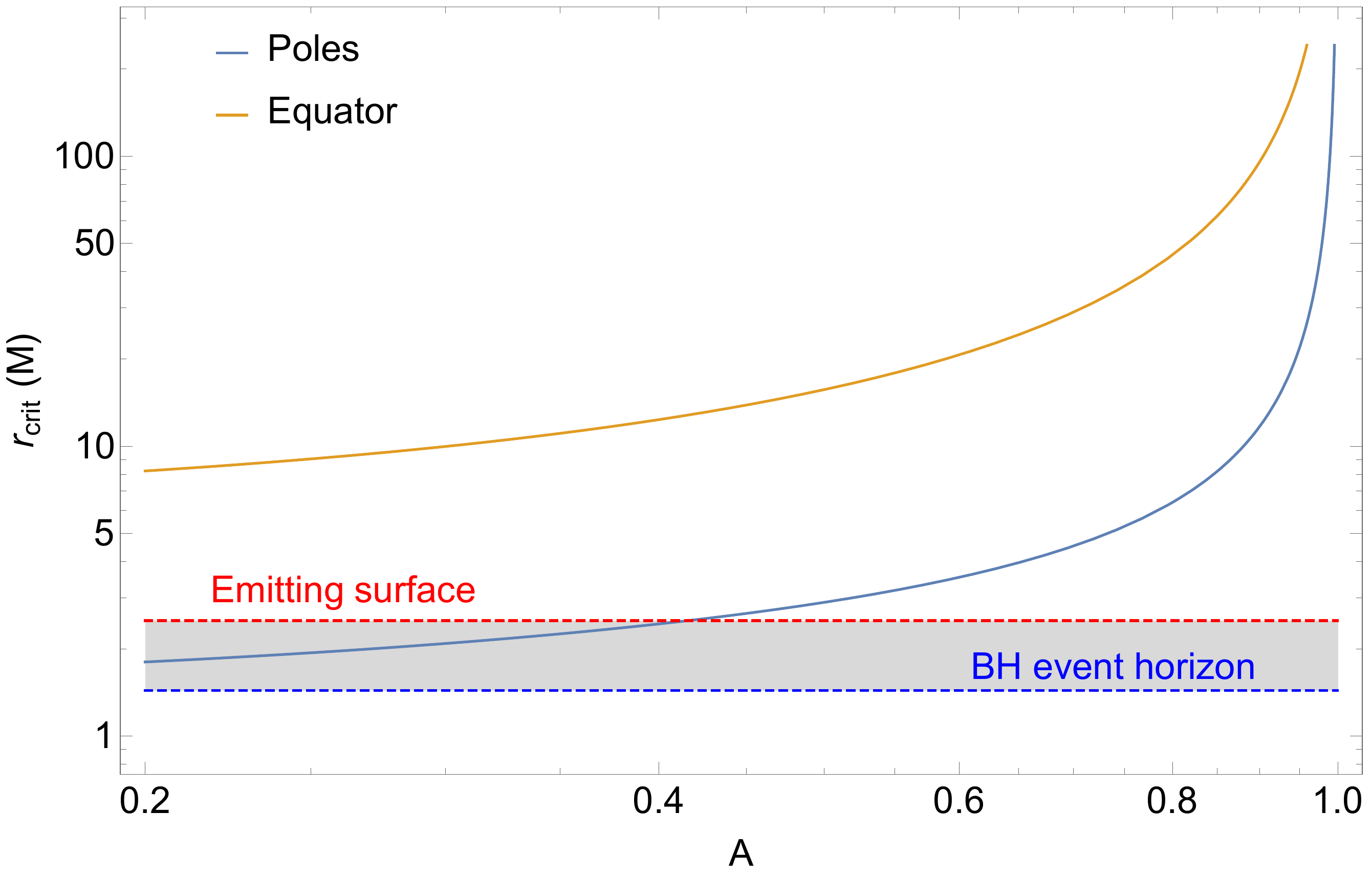}}
	\caption{Critical radius $r_{\rm (crit)}$ as a function of the luminosity parameter $A$ at the poles (blue line) and at the equator (orange line). The dashed red line represents the emitting surface, and the dashed blue line is the BH outer event horizon at $R_{\rm H}=1.44M$. In the left panel $f_\star=300$ Hz, while in the right panel $f_\star=1400$ Hz. The plot is constructed for the case of BH spin $a=0.9$ and the radius of the emitting surface $R_\star=2.5M$}
	\label{fig:Fig5}
\end{figure*}
\begin{figure}[h!]
	\centering
		\includegraphics[trim=0.5cm 0cm 0cm 0cm,scale=0.35]{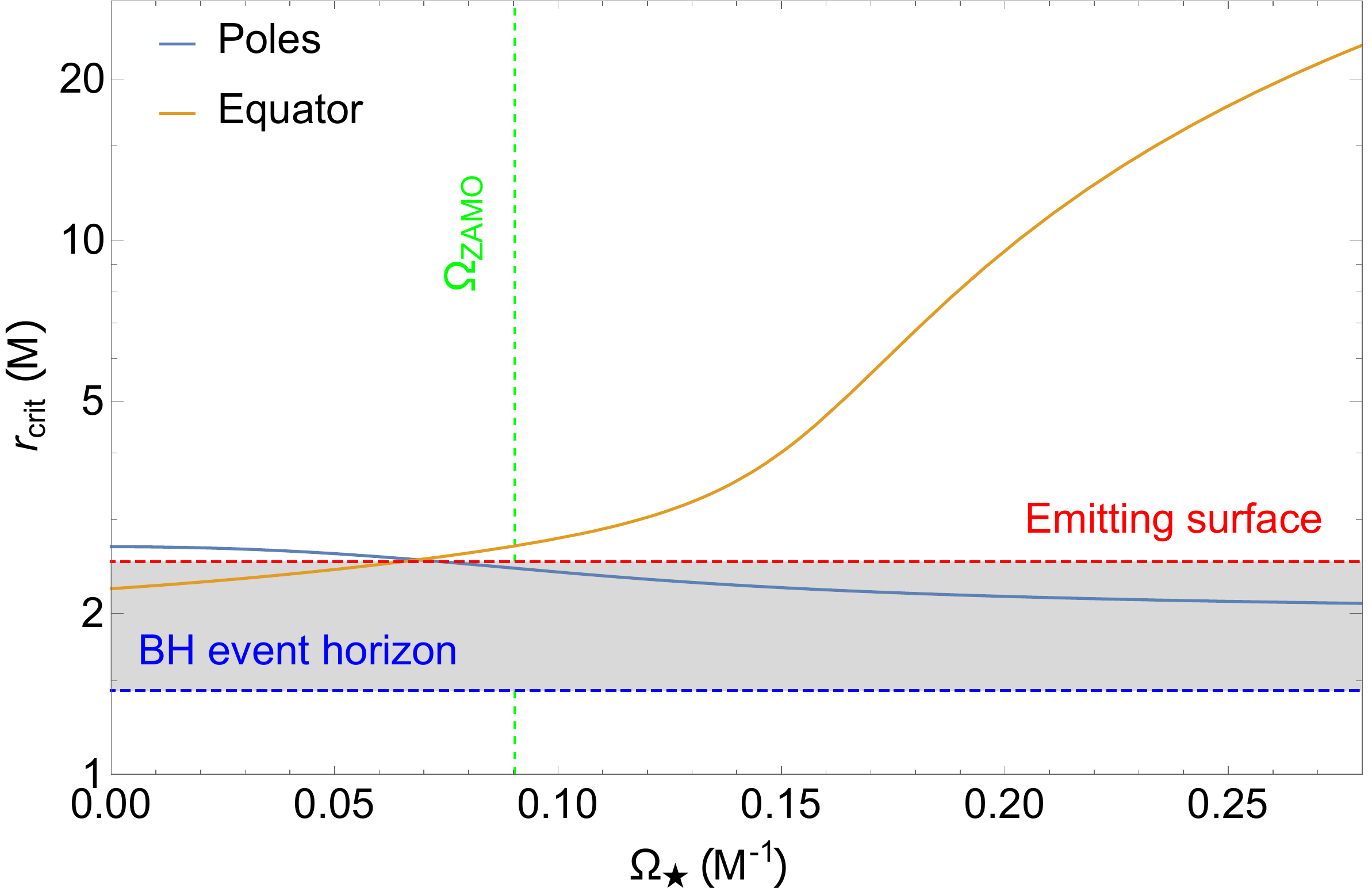}
	\caption{Critical radius $r_{\rm (crit)}$ as a function of the angular velocity of the emitting source $\Omega_\star$ at the poles (blue line) and at the equator (orange line). The dashed red line represents the emitting surface and the dashed blue line indicates the BH outer event horizon. The vertical dashed green line is the ZAMO angular velocity $\Omega_{\rm ZAMO}=0.09M^{-1}$ . The plot is constructed for a luminosity parameter of $A=0.5$, BH spin of $a=0.9$ and radius of the emitting surface of $R_\star=2.5M$}
	\label{fig:Fig6}
\end{figure}
\begin{figure}[h!]
	\centering
		\includegraphics[trim=0.5cm 0cm 0cm 0cm,scale=0.33]{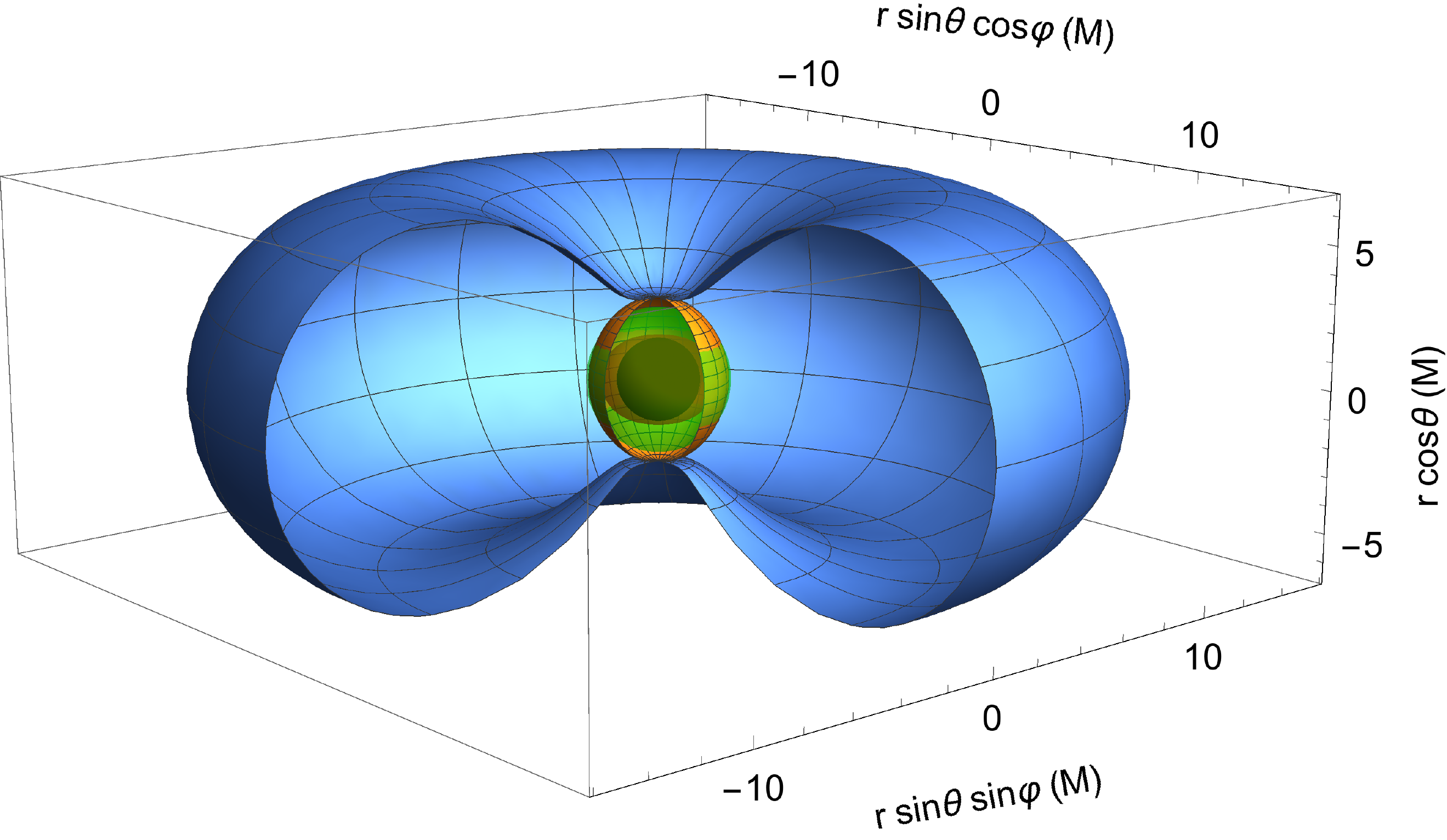}
	\caption{Critical hypersurfaces for a luminosity parameter of $A=0.5$, BH spin of $a=0.9$ and radius of the emitting surface of $R_\star=2.5M$. The black sphere is the BH outer event horizon with  size $R_{\rm H}=1.44M$, the red ellipsoid around the BH outer event horizon represents the ergosphere and the green sphere is the emitting surface. The orange surface is calculated for $\Omega_\star=0.05M^{-1}$ it has $r_{\rm (crit)}^{\rm eq}=2.42M$ and $r_{\rm (crit)}^{\rm pole}=2.89M$. The blue surface is obtained setting $\Omega_\star=0.24M^{-1}$ and has $r_{\rm (crit)}^{\rm eq}=15.65M$ and $r_{\rm (crit)}^{\rm pole}=2.89M$.}
	\label{fig:Fig7}
\end{figure}
\begin{figure*}[ht!]
	\centering
		\includegraphics[trim=0.5cm 0cm 0cm 0cm,scale=0.33]{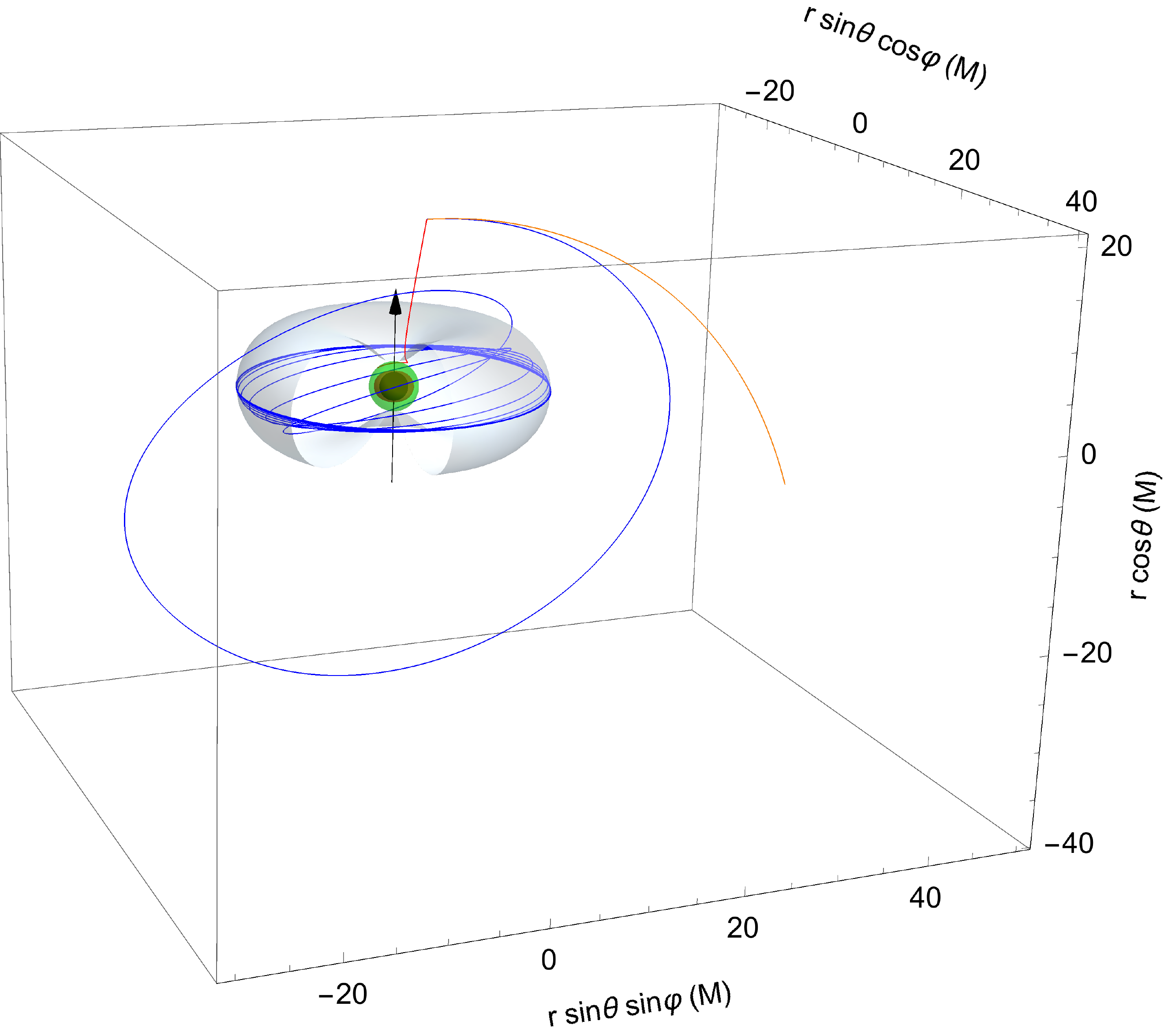}
		\hspace{0.3cm}
		\includegraphics[trim=0.5cm 0cm 0cm 0cm,scale=0.27]{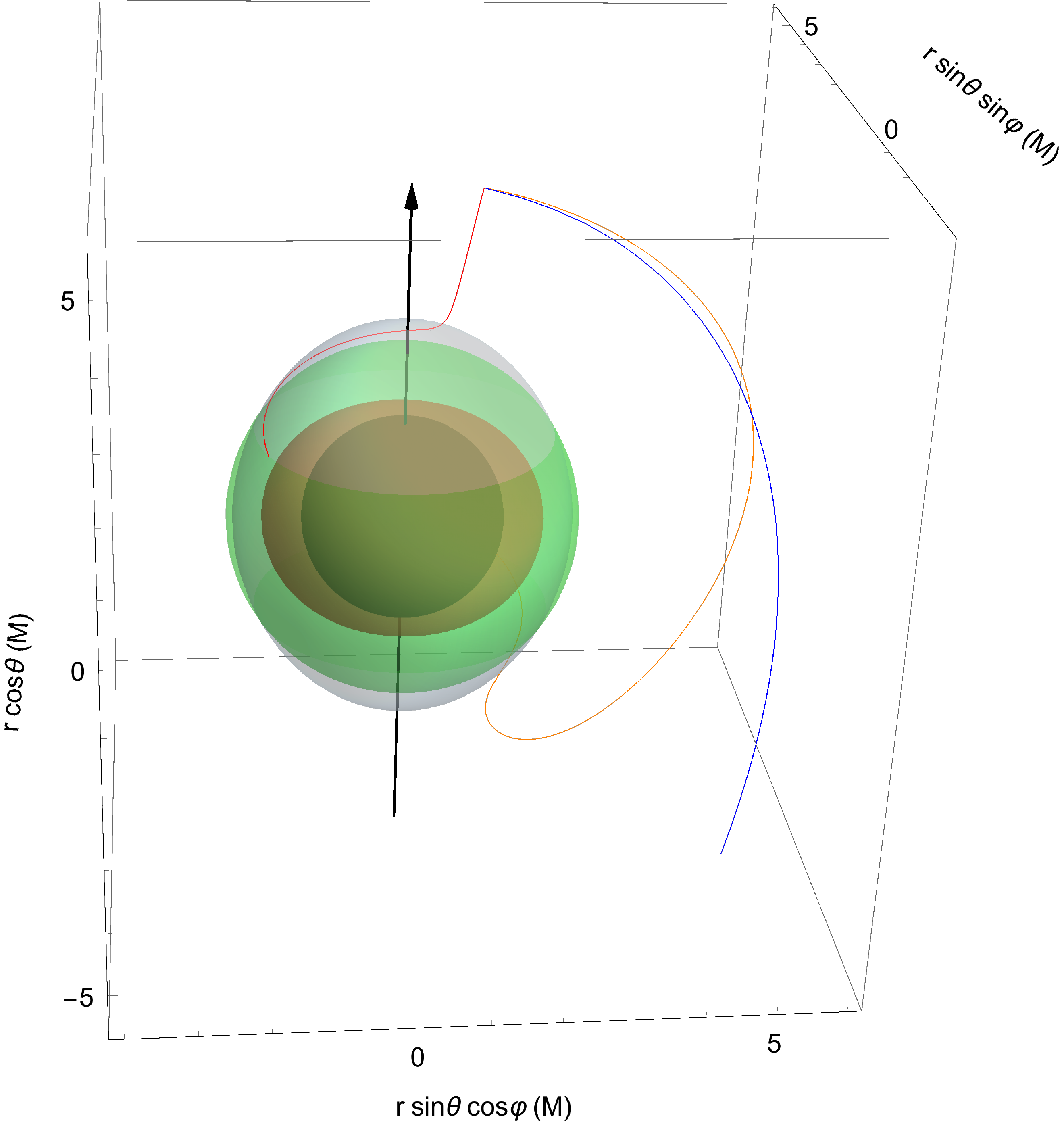}
	\caption{Test particle trajectories influenced by a radiation field with relative luminosity $A=0.5$ emitted from a surface at $R_\star=2.5M$ in the vicinity of BH with spin $a=0.9$. Left panel: the case of a rapidly rotating surface with $\Omega_\star\equiv 0.24M^{-1}$ ($f_\star=1550$ Hz). The test particles trajectories start outside the critical hypersurface at $r_0=20M,\,\theta_0=\pi/8$ with  initial velocity $\nu_0=0.22$ oriented in the  counterrotating azimuthal direction (blue), with $\nu_0=0.25$ oriented in the counterrotating azimuthal direction (orange) and with $\nu_0=0.22$ oriented radially towards the emitting surface (red). Right panel: The case of a slowly rotating surface with $\Omega_\star\equiv 0.05M^{-1} <\Omega_{\mathrm{ZAMO}}$ ($f_\star=323$ Hz). Test particles are emmited above the critical hypersurface forming two polar lobes at $r_0=5M,\,\theta_0=\pi/8$ with the initial velocity $\nu_0=0.1$ (red), $\nu_0=0.5$ (orange) and  $\nu_0=0.8$ (blue) oriented in the  counterrotating  azimuthal direction. The black sphere is the BH outer event horizon with size $R_{\rm H}=1.44M$, while the red ellipsoid around the BH outer event horizon represents the ergosphere and the green sphere is the emitting surface.}
	\label{fig:Fig14}
\end{figure*}

Figure \ref{fig:Fig5} shows the critical radius changes as a function of the  the luminosity parameters for two different rotation frequencies of the emitting surface $f_\star=300$ Hz ($\Omega_\star=0.05M^{-1}$) and $f_\star=1400$ Hz ($\Omega_\star=0.22M^{-1}$). 
The slow rotation case corresponds to the negative value of test field impact parameter on the equator $b=-1.55$ while the fast rotation case corresponds to to the positive one $b=2.67$
In the case of slow rotation with $f_\star=300$ Hz, the change from  a prolate to an oblate hypersurface takes place when the luminosity parameter exceeds $A\approx0.86$. 
For the range of luminosity values $A \in[0.425,0.53]$, the critical hypersurface only partially emerges from the emitting surface and forms two lobes around poles of the emitting surface.
In the case of fast rotation with $f_\star=1400$ Hz, the behaviour is simpler. The critical hypersurface always has an oblate shape. For luminosities $A<0.4$, the critical hypersurface forms a lobe around equator of the emitting surface, for higher luminosities the entire critical hypersurface is located outside the emitting surface.

In Fig. \ref{fig:Fig6}, we illustrate the behaviour of the critical hypersurface as a function of the angular velocity of the emitting surface $\Omega_\star$ for fixed luminosity $A=0.5$ and fixed spin parameter $a=0.9$. The $\Omega_\star$ interval ranges between zero and limiting value $\Omega_+(R_\star,a,\pi/2)=0.28M^{-1}$ ($f_\star=1808$ Hz) .
In the angular velocity range $\Omega_\star<0.07M^{-1}$ ($f_\star=452$ Hz) , the critical hypersurface forms two lobes around the poles of the emitting surfaces. For $\Omega_\star=0.07M^{-1}$, the critical hypersurface coincides with the emitting surface. For $\Omega_\star>0.07M^{-1}$, the critical hypersurface is a lobe around the equator of the emitting surface. In Fig. \ref{fig:Fig7}, we illustrate such behaviour by 3D plot of the critical hypersurface shape for the slowly rotating case with $\Omega_\star\equiv 0.05M^{-1}$ ($f_\star=323$ Hz) and for the fast rotating case with $\Omega_\star\equiv 0.24M^{-1}$ ($f_\star=1550$ Hz). The figure also clearly illustrates, that both hypersurfaces coincide on the polar axis, where photons coming radially from rigidly rotating emitting surface have always zero angular momentum ($b=0$).  

\subsubsection{Examples of test particle orbits in the vicinity of the critical hypersurface around BH}

In Fig. \ref{fig:Fig14}, we illustrate the results of the integration of some selected trajectories of test particles influenced by the radiation field emitted by a slowly or rapidly rotating surface as in the case corresponding to Fig. \ref{fig:Fig7}, i.e. for the value of luminosity parameter $A=0.5$ and the emitting surface located at $R_\star=2.5M$ in the vicinity of a BH with spin $a=0.9$.  
In the left panel, when the emitting surface rotates rapidly with $\Omega_\star=0.24M^{-1}$, the azimuthal impact parameter of the photons is $b=3.03$ on the equator. The test particles are emitted above the critical hypersurface at $r_0=20M,\,\theta_0=\pi/8$.  The particle emitted with initial velocity $\nu_0=0.22$ oriented in the azimuthal counter-rotating direction (blue) intersects the critical hypersurface and makes several loops inside it before it stabilizes on the circular equatorial orbit on the critical hypersurface. The particle emitted with initial velocity $\nu_0=0.25$ oriented also in the azimuthal counter-rotating direction (orange) escapes to infinity. Finally, the test particle emitted with an initial velocity  $\nu_0=0.22$ oriented radially towards the BH (red) falls on the emitting surface. 
On the right panel, when the emitting surface rotates slowly with $\Omega_\star\equiv 0.05M^{-1}<\Omega_{\mathrm{ZAMO}}$ ($f_\star=323$ Hz),  the azimuthal impact parameter of the photons is characterised  by a negative value of $b=-1.24$ on the equator. The test particles are emitted again above the critical hypersurface (which in this case consists of two polar lobe) at $r_0=5M,\,\theta_0=\pi/8$ with the velocity oriented in the azimuthal counter-rotating direction.  The particle emitted with initial velocity $\nu_0=0.1$ (red) is captured on the northern polar lobe, and after the latitudinal drift, it impacts the emitting surface at its intersection ring with the northern lobe.  Similarly the particle emitted with initial velocity $\nu_0=0.5$ (orange) after making a loop is captured on the southern polar lobe, and after the latitudinal drift, it impacts the emitting surface at its intersection ring with the southern lobe. Finally, the test particle emitted with initial velocity $\nu_0=0.8$ (blue) escapes to infinity. 

\section{Conclusions}
\label{sec:end}

We extended the general relativistic model of the 3D PR effect presented in \cite{DeFalco20183D} by 
considering the radiation field emitted in the purely radial direction in a local corotating frame from a rigidly rotating spherical source in the background of the Kerr spacetime geometry. This setup, though 
clearly idealised, may approximate the radiating surface of a rotating NS as well as that of a hot rotating corona in close vicinity of a BH. Owing to the rigid rotation of the emitting surface, the photon azimuthal impact parameter of the radiation field $b$ (a proxy of its angular momentum)  is a function of the polar angle $\theta$, with the highest value in the equatorial plane and zero value on the poles. In order to simplify the integration of test particles trajectories, we assume that the emitted photons are not moving in the latitudinal direction and their $\theta$ coordinate is conserved along the photon trajectory. Such a setup for the test radiation field is parametrized by the radius of the emitting surface $R_\star$ and its angular velocity $\Omega_\star$ and the emitted luminosity $A$. We have considered the range of possible values of azimuthal impact parameter of the radiation field $b$ depending on the parameters $R_\star$, $\Omega_\star$ and the spin of a NS or BH, when the maximum values of the angular velocity of the emitting surface $\Omega_\star$ are limited below superluminal rotation. In the case of a non-zero spin, the photon azimuthal impact parameter of the radiation field $b$ can attain both positive and negative values for positive (i.e. corotating) angular velocities $\Omega_\star$. In the case in which the emitting surface is partially located in the ergosphere and  rotates slowly with the angular velocity going to $\Omega_{\infty}(\RS)\equiv-\mathrm{g_{tt}}/\mathrm{g_{t\phi}}<\Omega_{\mathrm{ZAMO}}$, the azimuthal impact parameter of the radiation field $b$ diverges to $- \infty$. For extremely slowly rotating surfaces located in the ergosphere with $\Omega_{\star} <\Omega_{\infty}(\RS)$ $b$ is positive again. However, the photons emitted from such surfaces cannot escape the ergosphere. (see Fig. \ref{fig:lambdarange}).  

Using the \emph{observer-splitting formalism}, we formulated the equations of motion (\ref{EoM1})--(\ref{EoM6}) for test particles influenced by our test radiation field.
Their specific class of solutions corresponds to the axially-symmetric critical hypersurface, which is formed by test particles on the generally off-equatorial circular orbits around the emitting surface stabilized by the balance between gravitational attraction, radiation forces and PR drag. We have demonstrated that the shape of the critical hypersurface depends on the spin of the NS or BH and the parameters of the radiation field $R_\star$ and $\Omega_\star$. Depending on the interplay of such parameters, the critical hypersurface may morph between the oblate and prolate shape.

By using a cubic approximation of Eq. (\ref{eq:crit_hyper2}) to determine the location of critical hypersurface,  we found a criterion for distinguishing the regime with only one critical hypersurface  from the regime with three critical hypersurfaces. It should be noted, however, that the inner solutions of Eq. (\ref{eq:crit_hyper2}) are found in the close vicinity of the outer event horizon and in the region of spherical photon orbits, such that their physical relevance is likely very limited. Therefore, in the examples we analysed, multiple critical hypersurfaces are not relevant.
We determined the conditions for the  existence of off-equatorial suspended circular orbits bound on the critical hypersurface and determined the corresponding value of the polar angle $\psi$ of the velocity of the test particle $\nu$ as measured in the ZAMO frame as a function of latitudinal coordinate $\theta$. 

We analyzed in detail representative cases for both a rotating NS and a rotating BH.
In the case of a NS, its spherical surface is the source of the radiation field and the kerr parameter $a$ is proportional to the angular velocity of the emitting surface $\Omega_\star$; we found that the critical hypersurface always takes an oblate shape. Interestingly, over a limited range of relative luminosity values $A$, the critical hypersurface forms only a lobe around the NS equator. The angular velocity of the particles captured on such a hypersurface is equal to $\Omega_\star$ in the ring of intersection with the NS surface and decreases with the polar angle $\theta$ reaching a minimum on the equatorial plane (see Figs. \ref{fig:Fig4bis} and \ref{fig:Fig4bis1} )

In the case of an emitting surface approximating a spherical rotating hot-corona in the close vicinity of a BH, the critical hypersurface can take a prolate as well as an oblate shape. We analyzed the case of the emitting surface located outside the ergosphere of a BH where suspended off-equatorial orbits can exist. For high values of $\Omega_\star$, similar to the case of a NS, the critical hypersurface always takes an oblate shape, and over a range of relative luminosities $A$, it may form only a lobe around the equator of the emitting surface. In the case of slowly rotating emitting surfaces, an inversion effect may arise, where the prolate critical hypersurface forms two lobes around the poles of the emitting surface.  
In both cases, we integrated selected trajectories of test particles influenced by the interaction with the radiating field emitted from slowly as well as rapidly rotating emitting surfaces. 

A more complete analysis and classification of orbits in the radiation field emitted by a rigidly rotating spherical source, including the analysis of the stability of off-equatorial suspended orbits on the critical hypersurface and the conditions for capturing the test particles on such orbits will be the subject of a separate study.

\section*{Acknowledgements}
P.B., K.G. and D.L. acknowledge the Czech Science Foundation (GAČR) grant GAČR 17-16287S and internal grant of Silesian University in Opava SGS/13/2019.
V.D.F. and E.B. thank the Silesian University in Opava for having funded this work. P.B., V.D.F. and E.B. thank the Osservatorio Astronomico di Roma in Monteporzio Catone for the hospitality. V.D.F. and E.B are grateful to Gruppo Nazionale di Fisica Matematica of Istituto Nazionale di Alta Matematica for support. LS acknowledges financial contributions from ASI-INAF agreements 2017-14-H.O and  I/037/12/0 and from ``iPeska'' research grant (P.I. Andrea Possenti) funded under the INAF call PRIN-SKA/CTA (resolution 70/2016).

\begin{appendix}
\section{Classical 3D limit with nonzero photon impact parameter}
\label{Appendix_Classic_Limit}
In this section we report the classical limit of Eqs. (\ref{EoM1}) -- (\ref{EoM3}). We already know how to treat the kinematic part from our previous study \cite{DeFalco20183D}, so we focus our attention on the radiation force $\mathcal{F}_{\rm (rad)}(U)^{\alpha}$ (\ref{Frad0}). We first consider the Schwarzschild limit $a=0$ of Eq. (\ref{eq: sigma_tilde}), for which we obtain 
\begin{equation}
\begin{split}
& \sigma \left[\Phi E(U) \right]^2 =\\
& \frac{A \gamma^2 \left[ 1 - \nu \sin \psi \cos \left( \alpha-\beta\right) \right]^2}{r \left(1-\dfrac{2M}{r}\right) \sqrt{r^2- \left(1 - \dfrac{2M}{r}\right)\left(q+b^2\right)}}.
\end{split}
\end{equation}
Then for $r\rightarrow +\infty$, we have
\begin{equation}
\begin{split}
&\sigma \left[\Phi E(U) \right]^2 \approx \dfrac{A \gamma^2 \left[ 1 - \nu \sin \psi \cos \left( \alpha-\beta\right) \right]^2} {r^2}.
\end{split}
\end{equation}
Finally, we consider  $\nu \rightarrow 0$; the radiative force components Eqs. (\ref{rad1})--(\ref{rad3}) reduce to
\begin{eqnarray}
&&\mathcal{F}_{\rm (rad)}(U)^{\hat{r}}= \dfrac{A}{r^2} \left[ \sin \beta \right.\nonumber\\
&&\left.\hspace{2cm}-\dot{r}(1+\sin^2 \beta) -r\dot{\varphi}\sin\theta\frac{\sin(2\beta)}{2}\right],\label{App_1}\\
&&\mathcal{F}_{\rm (rad)}(U)^{\hat{\theta}}=-\dfrac{A}{r} \dot{\theta},\label{App_2}\\
&&\mathcal{F}_{\rm (rad)}(U)^{\hat{\varphi}}=\dfrac{A}{r^2} \left[ \cos \beta\right.\nonumber\\ 
&&\left.\hspace{1.8cm}-\dot{r}\frac{\sin(2\beta)}{2}-r\dot{\varphi} \sin \theta \left(1 +\cos^2 \beta\right) \right],\label{App_3}
\end{eqnarray}
where we exploited the following approximations of the test particle velocity components (cf. Eqs. (\ref{four_velocity}))
\begin{eqnarray}
U^r &&\equiv\dot{r}\approx\nu\sin\psi\sin\alpha,\label{App_vel_U1} \\
U^\theta &&\equiv \dot{\theta}\approx\frac{\nu\cos\psi}{r},\label{App_vel_U2}\\
U^\varphi &&\equiv \dot{\varphi}\approx\frac{\nu\sin\psi\cos\alpha}{r\sin\theta},\label{App_vel_U3}
\end{eqnarray}
here the dot indicates the derivative with respect to the time $t$. It should be stressed that the classical 3D result concerning the radial radiation field (reported in the Appendix of Ref. \cite{DeFalco20183D}) is recovered easily  for $\beta=\pi/2$ in Eqs. (\ref{App_1})--(\ref{App_3}). Therefore the related classical equations of motion read as
\begin{eqnarray}
&&\ddot{r}-r\dot{\varphi}^2\sin^2\theta-r\dot{\theta}^2+\frac{GM}{r^2}=\frac{Ac}{r^2}\sin\beta\nonumber\\
&&-\frac{A}{r^2}\left[\dot{r}(1+\sin^2 \beta)+r\dot{\varphi}\sin\theta\frac{\sin(2\beta)}{2}\right],\label{eqm1}\\
&&r\ddot{\theta}+2\dot{r}\dot{\theta}-r\dot{\varphi}^2\sin\theta\cos\theta=-A\frac{\dot{\theta}}{r}\label{eqm2},\\
&&r\ddot{\varphi}\sin\theta+2\dot{r}\dot{\varphi}\sin\theta+2r\dot{\theta}\dot{\varphi}\cos\theta=\frac{A}{r^2}\cos \beta \nonumber\\
&&-\frac{A}{r^2} \left[\dot{r}\frac{\sin(2\beta)}{2}+r\dot{\varphi} \sin \theta \left(1 +\cos^2 \beta\right)\right],\label{eqm3}
\end{eqnarray}
where on the right-hand side of Eqs. (\ref{eqm1}) and (\ref{eqm3}) we have the radiation pressure projected along the radial and azimuthal direction of the emitted photon and the PR effect is multiplied by a factor taking into account the non-radial direction of the emitted photon. In addition we note that the non-radial emission of the photons breaks the spherical symmetry of the equations of motion (see Appendix of Ref. \cite{DeFalco20183D}). Classically the azimuthal photon angle $\beta$ is related to the photon impact parameter $b$ through (see Eq. (\ref{ANG1}))
\begin{equation}
\cos\beta=\frac{b}{r\sin\theta}.
\end{equation}
In our previous work \cite{DeFalco20183D}, we saw that the time component of the equations of motion corresponds to the energy balance 
\begin{equation} 
\begin{aligned}
&\frac{d}{dt}\left(\frac{\nu^2}{2}-\frac{GM}{r}\right)=\frac{Ac}{r^2}\left(\dot{r}\sin\beta+r\dot{\varphi}\sin\theta\cos\beta\right)\\
&-\frac{A}{r^2}\left[\dot{r}^2(1+\sin^2 \beta)+r^2\dot{\theta}^2\right.\\
&\left.+r^2\dot{\varphi}^2 \sin^2 \theta \left(1 +\cos^2 \beta\right)+\dot{\varphi}\dot{r}r\sin\theta\sin(2\beta)\right],
\end{aligned}
\end{equation}
where the left term represents the total mechanical energy, while the right term corresponds to the radiation pressure and the energy dissipated through the PR effect.

\section{Weak field limit with slow and fast rotations}
\label{Appendix_Weak_Field}
We determine here the weak field approximation ($r\to\infty$) of the equations of motion (\ref{EoM1})--(\ref{EoM3}), at the first order in the spin parameter $a$ for slow rotations ($a \to 0$). Applying such limits to Eq. (\ref{eq: sigma_tilde}) we get
\begin{equation}
\sigma \left[\Phi E(U) \right]^2 \approx A\ \gamma^2\ \Gamma(\nu)\ f(a,r)
\end{equation}
where $\Gamma(\nu) = \left[ 1 - \nu \sin \psi \cos \left( \alpha-\beta\right) \right]^2$ and 
\begin{equation}
\label{eq: f:function}
\begin{split}
& f(a,r) = \dfrac{1}{r^2}+\dfrac{2M}{r^3} +\dfrac{8M^2+b^2+q}{2r^4}+ \dfrac{8M^3}{r^5}\\
&\hspace{1cm}-\dfrac{2Mb}{r^5}a+{\rm O}\left(\frac{1}{r^6}\right)+{\rm O}\left(\frac{a}{r^6}\right) +{\rm O}\left(\frac{a^2}{r^4}\right).
\end{split}
\end{equation}
Therefore, the radiation field components read as
\begin{eqnarray}
&&\mathcal{F}_{\rm (rad)}(U)^{\hat{r}}\approx A\ \gamma^2\ \Gamma(\nu)\ f(a,r)\ \mathcal{V}^{\hat r}, \label{WFApp_1}\\
&&\mathcal{F}_{\rm (rad)}(U)^{\hat{\theta}}\approx A\ \gamma^2\ \Gamma(\nu)\ f(a,r)\ \mathcal{V}^{\hat \theta}, \label{WFApp_2}\\
&&\mathcal{F}_{\rm (rad)}(U)^{\hat{\varphi}}\approx A\ \gamma^2\ \Gamma(\nu)\ f(a,r)\ \mathcal{V}^{\hat \varphi}.\label{WFApp_3}
\end{eqnarray}
For slow motion ($\nu \rightarrow 0$), Eqs. (\ref{WFApp_1})--(\ref{WFApp_3}) reduce to 
\begin{eqnarray}
&&\mathcal{F}_{\rm (rad)}(U)^{\hat{r}}\approx A\ f(a,r) \left[ \sin \beta-\dot{r}(1+\sin^2 \beta) \right. \nonumber\\
&&\left.\hspace{2cm}-\left(r\dot{\varphi}-\frac{2aM}{r^2}\right)\sin\theta\frac{\sin(2\beta)}{2}\right],\label{App_4}\\
&&\mathcal{F}_{\rm (rad)}(U)^{\hat{\theta}}\approx-A\ f(a,r)\  r\dot{\theta}, \label{App_5}\\
&& \mathcal{F}_{\rm (rad)}(U)^{\hat{\varphi}}\approx A\ f(a,r) \left[\cos \beta-\dot{r}\frac{\sin(2\beta)}{2} \right. \nonumber\\
&&\left.\hspace{1.6cm} -\left(r\dot{\varphi}-\frac{2aM}{r^2}\right)\sin\theta \left(1 +\cos^2 \beta\right)\right],\label{App_6}
\end{eqnarray}
where we have exploited the following approximations for the test particle velocity components
\begin{eqnarray}
U^r &&\equiv\dot{r}\approx\nu\sin\psi\sin\alpha+{\rm O}\left(\frac{1}{r}\right)+{\rm O}\left(\frac{a^2}{r^2}\right),\label{App_new_velocity1}\\
U^\theta &&\equiv \dot{\theta}\approx\frac{\nu\cos\psi}{r}+{\rm O}\left(\frac{1}{r^3}\right)+{\rm O}\left(\frac{a^2}{r^3}\right),\label{App_new_velocity2}\\
U^{\varphi} &&\equiv \dot{\varphi} \approx \dfrac{\nu \cos \alpha \sin \psi}{r \sin \theta}\nonumber\\
&&+\dfrac{2Ma}{r^3}+{\rm O}\left(\frac{1}{r^4}\right)+{\rm O}\left(\frac{a^2}{r^3}\right)+{\rm O}\left(\frac{a}{r^4}\right); \label{App_new_velocity3}
\end{eqnarray}
here dot means the derivative with respect to the affine parameter $\tau$. 
We note that Eqs. (\ref{App_new_velocity1})--(\ref{App_new_velocity2}) are similar to Eqs. (\ref{App_vel_U1})--(\ref{App_vel_U2}) in the classical limit. In addition, in Eqs. (\ref{App_4}) and (\ref{App_6}) by substituting the azimuthal test particle velocity Eq. (\ref{App_new_velocity3}), there is a factor $a$ that when multiplied by $f(a,r)$ gives a term of second order in $a$ that must be neglected. Therefore, we have
\begin{eqnarray}
&&\mathcal{F}_{\rm (rad)}(U)^{\hat{r}}\approx A\left\{\ f_1(r) \left[ \sin \beta-\dot{r}(1+\sin^2 \beta)\right]\right. \nonumber\\
&&\left.-f_2(a,r)\dot{\varphi}\sin\theta\frac{\sin(2\beta)}{2}-\frac{2aM}{r^4}\sin\theta\frac{\sin(2\beta)}{2}\right\}\nonumber\\
&&+{\rm O}\left(\frac{1}{r^5}\right)+{\rm O}\left(\frac{a^2}{r^3}\right)+{\rm O}\left(\frac{a}{r^5}\right),\label{App_42}\\
&&\mathcal{F}_{\rm (rad)}(U)^{\hat{\theta}}\approx-A\ f_2(a,r)\  r\dot{\theta}\nonumber\\
&&+{\rm O}\left(\frac{1}{r^5}\right)+{\rm O}\left(\frac{a^2}{r^3}\right)+{\rm O}\left(\frac{a}{r^5}\right), \label{App_52}\\
&& \mathcal{F}_{\rm (rad)}(U)^{\hat{\varphi}}\approx A\left\{f_1(r) \left[\cos \beta-\dot{r}\frac{\sin(2\beta)}{2} \right]\right. \nonumber\\
&&\left.-f_2(a,r)\dot{\varphi}\sin\theta \left(1 +\cos^2 \beta\right)-\frac{2aM}{r^4}\sin\theta \left(1 +\cos^2 \beta\right)\right\}\nonumber\\
&&+{\rm O}\left(\frac{1}{r^5}\right)+{\rm O}\left(\frac{a^2}{r^3}\right)+{\rm O}\left(\frac{a}{r^5}\right),\label{App_62}
\end{eqnarray}
where
\begin{eqnarray}
f_1(r)&&=\dfrac{1}{r^2}+\dfrac{2M}{r^3} +\dfrac{8M^2+b^2+q}{2r^4},\label{fun1}\\
f_2(a,r)&&=\dfrac{1}{r}+\dfrac{2M}{r^2} +\dfrac{8M^2+b^2+q}{2r^3}\nonumber\\
&&+ \dfrac{8M^3}{r^4}-\dfrac{2Mb}{r^4}a.\label{func2}
\end{eqnarray}
We note that the impact parameters appear already at the $r^{-3}$-order, whereas the spin parameter appear for the first time at the $r^{-4}$ term. \emph{The slow-rotation configures as an effect of the fourth-order in the general relativistic radiation processes.} Note also that the next order in the spin parameter is still linear, whereas the second order appears at the third order. 

Now we have also to consider the approximation of the geometric part (see Tab. I in Ref. \cite{DeFalco20183D}, for further details), which yields
\begin{eqnarray}
a(U)^{\hat{r}}&&= \frac{d}{d\tau}\left( \gamma\dot{r} \right)-\frac{6\gamma^2 aM\dot{\varphi}\sin\theta}{r^2}\nonumber\\
&&-\frac{\gamma^2}{r}\left[r^2\dot{\theta}^2+\left(r^2\dot{\varphi}^2-\frac{4Ma\dot{\varphi}}{r}\right)\sin^2\theta\right]\nonumber\\
&&+\frac{M\gamma^2}{r^2}\left[1+r^2\dot{\theta}^2+r^2\sin^2\theta\right]\nonumber\\
&&+\frac{M\gamma^2}{r^3}\left[1+\frac{1}{2}\left(r^2\dot{\theta}^2+r^2\sin^2\theta\right)\right],\nonumber\\
&&+ {\rm O}\left(\frac{1}{r^4}\right)+{\rm O}\left(\frac{a}{r^3}\right)+ {\rm O}\left(\frac{a^2}{r^3}\right)\label{GP_1}\\
a(U)^{\hat{\theta}}&&= \frac{d}{d\tau} \left(\gamma r\dot{\theta}\right)-\frac{\gamma^2 M\dot{r}\dot{\theta}}{r}+\frac{\gamma^2}{r}\left[r\dot{\theta}\dot{r}-\left(r\dot{\varphi}-\frac{4aM\dot{\varphi}}{r}\right)\right],\nonumber\\
&& + {\rm O}\left(\frac{1}{r^3}\right)+{\rm O}\left(\frac{a}{r^3}\right)+ {\rm O}\left(\frac{a^2}{r^3}\right),\label{GP_2}\\
a(U)^{\hat{\varphi}}&&= \frac{d}{d\tau}\left[\gamma\left(r\dot{\varphi}-\frac{2Ma}{r^2}\right)\right]-\frac{\gamma^2 M\dot{r}\dot{\varphi}}{r},\nonumber\\
&&\frac{\gamma^2}{r}\left[\left(r\dot{\varphi}-\frac{2Ma}{r^2}\right)\left(r\dot{\theta}\cos\theta+\dot{r}\sin\theta\right)\right],\nonumber\\
&&+{\rm O}\left(\frac{1}{r^4}\right)+{\rm O}\left(\frac{a}{r^4}\right)+ {\rm O}\left(\frac{a^2}{r^3}\right).\label{GP_3}
\end{eqnarray}
\emph{In the geometric part the spin appears linearly already at the third order.} This means that the linearized frame dragging effect has more relevance in the gravitational-geometric part than in the radiation processes. The second order effect in the spin appear also at the third order. In this case, it is important to note that the affine parameter $\tau$ is related to the coordinate time $t$ not linearly as in the classic limit, but it satisfies this condition (see $U^t$ in Eq. \ref{four_velocity})
\begin{equation}\label{App_new_velocity4}
U^t \equiv\dot{t}\approx \gamma\left(1+\frac{M}{r}\right)+{\rm O}\left(\frac{1}{r^2}\right)+{\rm O}\left(\frac{a^2}{r^3}\right),
\end{equation}
and in the slow motion approximation, it reduces to
\begin{equation}\label{App_new_velocity42}
U^t \equiv\dot{t}\approx\left(1+\frac{M}{r}\right).
\end{equation}
Therefore, in order to pass from the proper time $\tau$ derivative to the coordinate time $t$ derivative, we have to multiply such $t$-derivative by the geometrical factor of Eq. (\ref{App_new_velocity42}). In the slow-motion limit the test particle acceleration components Eqs. (\ref{GP_1})--(\ref{GP_3}) become
\begin{eqnarray}
a(U)^{\hat{r}} &&\approx \ddot{r} +\dfrac{2aM\dot{\varphi}\sin\theta(2\sin\theta-3)}{r^2}\nonumber\\
&&+\frac{M}{r^2}\left(1+\frac{M}{r}\right),\label{App_9}\\
a(U)^{\hat{\theta}} &&\approx \dot{r}\theta+r\dot{\theta}+\frac{4aM\dot{\varphi}\cos\theta}{r^2},\label{App_9a}\\
a(U)^{\hat{\varphi}} &&\approx r\ddot{\varphi}\sin\theta+\dot{r}\dot{\varphi}\sin\theta+r\dot{\varphi}\dot{\theta}\cos\theta\nonumber\\
&&-\frac{2Ma}{r^3}\left(r\dot{\theta}\cos\theta+\dot{r}\sin\theta\right).\label{App_9b}
\end{eqnarray}

\end{appendix}

\bibliography{references}

\end{document}